\documentclass[pdflatex,sn-mathphys-num]{sn-jnl}

\usepackage{graphicx}%
\usepackage{multirow}%
\usepackage{amsmath,amssymb,amsfonts}%
\usepackage{amsthm}%
\usepackage{mathrsfs}%
\usepackage[title]{appendix}%
\usepackage{xcolor}%
\usepackage{textcomp}%
\usepackage{manyfoot}%
\usepackage{booktabs}%
\usepackage{algorithm}%
\usepackage{algorithmicx}%
\usepackage{algpseudocode}%
\usepackage{listings}%
\usepackage{bm}
\theoremstyle{thmstyleone}%
\theoremstyle{thmstyletwo}%

\theoremstyle{thmstylethree}%

\begin{document}

\title[Article Title]{Topological Lithography via External Field: Creating Topological States Anywhere Beyond the Edge}


\author[1]{\fnm{Haoran} \sur{Nie}}\email{haorannie@link.cuhk.edu.hk}

\author[1]{\fnm{Chaoran} \sur{Jiang}}\email{chaoranjiang@cuhk.edu.hk}

\author*[2]{\fnm{Xiangying} \sur{Shen}}\email{shenxy66@sysu.edu.cn}

\author*[1,3]{\fnm{Lei} \sur{Xu}}\email{xuleixu@cuhk.edu.hk}

\affil[1]{\orgdiv{Department of Physics}, \orgname{The Chinese University of Hong Kong}, \orgaddress{ \country{Hong Kong SAR}}}

\affil[2]{\orgdiv{School of Science}, \orgname{Shenzhen Campus of Sun Yat-Sen University}, \orgaddress{\city{Shenzhen}, \country{China}}}

\affil[3]{\orgdiv{Shenzhen Research Institute}, \orgname{The Chinese University of Hong Kong}, \orgaddress{ \city{Shenzhen}, \country{China}}}


\abstract{Topological insulators (TIs), recognized for their robust boundary states and unconventional phase transitions, have emerged as one of the most impactful discoveries in recent decades, attracting considerable interest across diverse fields. However, conventional TIs require topological contrasts between adjacent bulk regions, typically achieved by distinct symmetries, which limits their flexibility and broader applicability. In this work, we introduce a “lithography” approach to inducing topological states by applying external fields in time-reversal-symmetric systems. These states transcend the conventional bulk topology design paradigm and offer exceptional tunability: they can exist at geometric edges, within the bulk, or even be induced remotely. Because the external field is highly controllable, the induced states are programmable, reconfigurable, and can be easily tailored into desirable patterns. Theoretically, we demonstrate that the external field modifies the Jackiw-Rebbi mechanism, inducing a real-space topological transition characterized by a local topological marker (LTM). We further establish an extended valley bulk-edge correspondence, which explains both the conventional scenario and our findings. Our results, validated across mechanical, electronic, and acoustic platforms, highlight the broad applicability of this approach to various systems. This work not only advances the theory of topology but also enhances the diversity, tunability, and practical implementation of topological states and materials.}

\keywords{valleytronics, topological insulator, bulk-edge correspondence, lithography}



\maketitle

\section*{Introduction}\label{sec1}

Topological insulators (TIs) have emerged as one of the most captivating research areas in recent years, deepening our understanding of material phases in wave systems \cite{restaInsulatingStateMatter2011,hastingsTopologicalInsulatorsCalgebras2011}. Topological states are protected by the system's global topology and remain robust as long as the energy gap is not closed, enabling applications in low-loss waveguides, high-precision sensing, next-generation chips, and quantum information processing. Recently, topological materials have become closely intertwined with prominent research fields, including non-Hermitian physics \cite{ashidaNonHermitianPhysics2020,songNonHermitianTopologicalInvariants2019,zhaoNonHermitianTopologicalLight2019,linObservationTopologicalTransition2024}, nonlinear systems \cite{maNonlinearTopologicalMechanics2023,tuloupNonlinearityInducedTopological2020,soneNonlinearityinducedTopologicalPhase2024,loTopologyNonlinearMechanical2021,zhouTopologicalInvariantAnomalous2022a}, non-reciprocal interactions \cite{scheibnerOddElasticity2020a,coulaisStaticNonreciprocityMechanical2017}, and Moiré lattices \cite{oudichEngineeredMoirePhotonic2024,dongFlatBandsMagicAngle2021,rosendolopezFlatBandsMagicAngle2020,xuHydrodynamicMoireSuperlattice2024,zhaoRealizationHaldaneChern2024,alezziTopologicalFlatBands2024}, making it a dynamic and multidisciplinary research focus. This field has progressed beyond electronic quantum systems \cite{restaInsulatingStateMatter2011,chiuClassificationTopologicalQuantum2016,yanTopologicalMaterialsWeyl2017} to include photonic \cite{barikTopologicalQuantumOptics2018,chaudronElectricfieldinducedMultiferroicTopological2024,rechtsmanPhotonicFloquetTopological2013,fangRealizingEffectiveMagnetic2012,wuSchemeAchievingTopological2015,hafeziImagingTopologicalEdge2013,caceres-aravenaCompactTopologicalEdge2024}, acoustic \cite{yangTopologicalAcoustics2015,xueAcousticHigherorderTopological2019,yangAcousticTypeIIWeyl2016,chenAcousticWeylPoints2018,xiaoSyntheticGaugeFlux2015a,laiTopologicalPhononicFiber2024}, fluid, thermal \cite{xuObservationBulkQuadrupole2023,liLocalizedDelocalizedTopological2024}, and elastic systems \cite{huberTopologicalMechanics2016,zhaoElasticValleySpin2022,chenElasticQuantumSpin2018,fanElasticHigherOrderTopological2019,shiDisorderinducedTopologicalPhase2021,nashTopologicalMechanicsGyroscopic2015,kaneTopologicalBoundaryModes2014a,nie2025designing}. This evolution has also provided new perspectives for the development of metamaterials \cite{chenVariousTopologicalPhases2023a,pauloseTopologicalModesBound2015,yanPseudomagneticFieldsEnabled2021,zhangProgrammableElasticValley2019,maExperimentalDemonstrationDualBand2021a,cuiOnChipElasticWave2024}. The search for new topological phases and efforts to expand their practical applications have established topological materials as a promising and rapidly evolving research direction \cite{niObservationHigherorderTopological2019,wangStructuralAmorphizationInducedTopological2022,zhouTopologicalEdgeFloppy2018,yangStrainInducedGaugeField2017}.

In many topological systems, and particularly in valley-Hall platforms considered here, protected transport channels are realized at interfaces separating regions with distinct effective topological characteristics.\cite{schaibleyValleytronics2DMaterials2016,bernevigTopologicalInsulatorsTopological2013} Consequently, their spatial locations are usually encoded by the underlying material domains or lattice structure. Therefore, constructing topological insulators (TIs) requires controlling the global symmetry of bulk regions to generate the necessary topological differences. However, this process is often costly, inflexible, and difficult to reconfigure. For instance, in valley Hall insulators (VHIs), achieving topological states necessitates breaking the spatial inversion symmetry of the bulk regions. A common method involves swapping sublattice points between bulks \cite{schaibleyValleytronics2DMaterials2016,liuTunableAcousticValleyHall2018,zhangProgrammableElasticValley2019}, which entails assembling two bulk materials with inverted sublattice distributions—a process that is complex and hard to modify \cite{zhaoElasticValleySpin2022,liuTunableAcousticValleyHall2018}. Similarly, even in amorphous systems, topological states remain constrained by bulk symmetry \cite{wangStructuralAmorphizationInducedTopological2022,shiTopologicalPhaseTransition2022,agarwalaTopologicalInsulatorsAmorphous2017}, as these states rely on edge modes formed at the amorphous bulk interface under specific symmetry manipulations \cite{nashTopologicalMechanicsGyroscopic2015,biancoMappingTopologicalOrder2011a}. Realizing such states often requires active control to break time-reversal symmetry, further increasing experimental complexity. As such, the development of flexible platforms that go beyond conventional bulk-engineering strategies remains an important and ongoing research challenge.

\begin{figure}[htbp]
\begin{center}
\centerline{\includegraphics[width=0.95\linewidth]{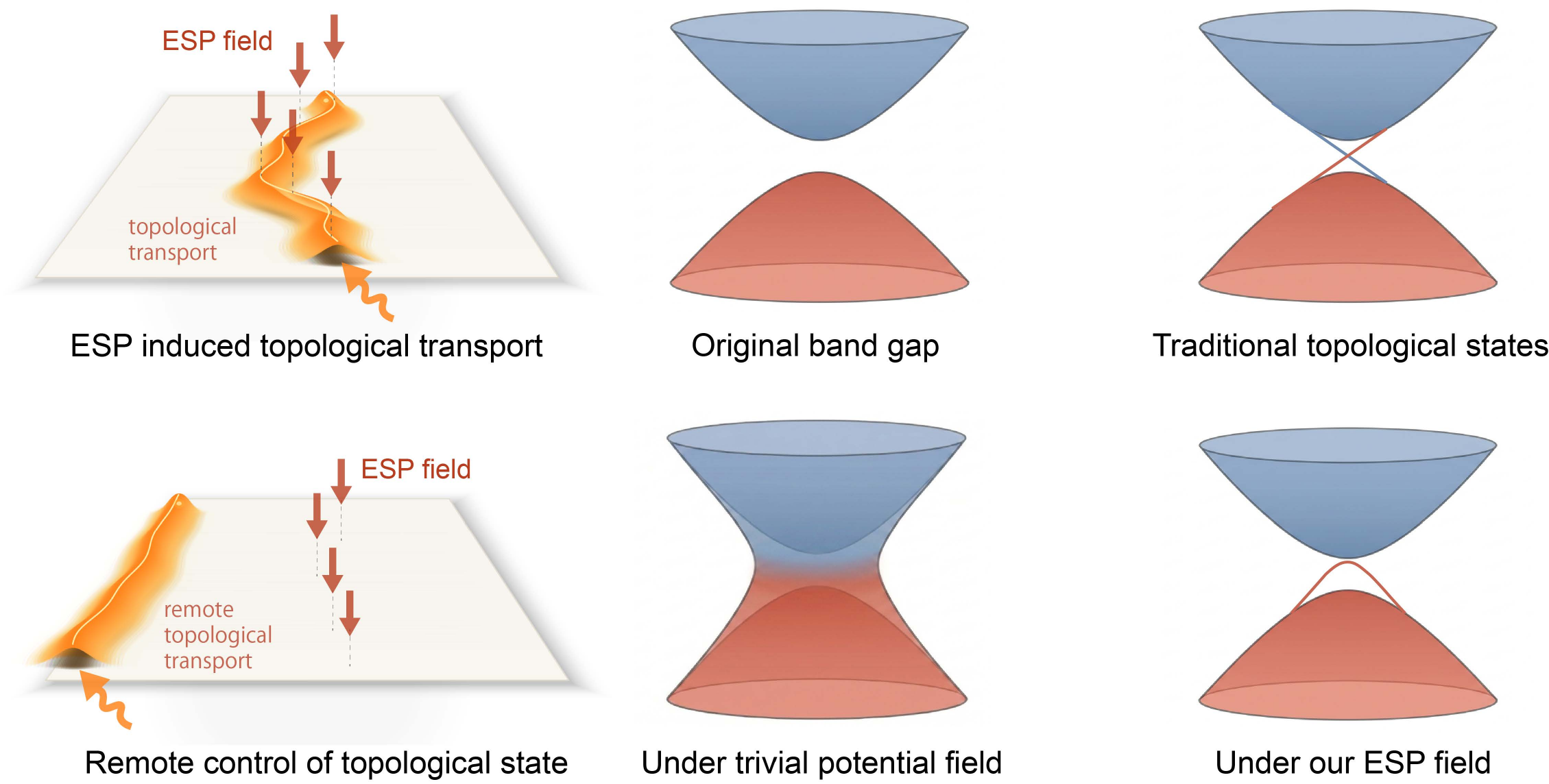}}
\caption{\label{fig.1} Schematic diagram of ESP-manipulated topological transport and band structure schematics of different scenarios. Traditional topological edge states are induced by changing bulk symmetry across the boundary. Trivial potential field may close the gap but destroy topological structure. Our ESP field reshapes the effective mass and induces a topological state without reconfiguring the lattice geometry or fabricating distinct bulk domains.}
\vspace{-0.5cm}
\end{center}
\end{figure}

In this study, we identify a field-driven route to creating and positioning topological states without fabricating distinct bulk domains. Instead, these topological states are induced by an external scalar potential (ESP) field, which alters the band structure and shifts the domain wall within the Jackiw–Rebbi framework. Unlike conventional approaches based on real or pseudo magnetic fields (i.e., vector potential) or doping \cite{cuiOnChipElasticWave2024,brendelPseudomagneticFieldsSound2017,yangStrainInducedGaugeField2017,quTopologicalPhotonicAlloy2024}, our mechanism employs a scalar potential such as an electric field instead of a magnetic field. Figure 1 illustrates the differences between the conventional scenarios and our proposed ESP (external scalar potential) mechanism through schematic diagrams.

Moreover, we demonstrate that these topological states are not confined to geometric boundaries but can be spatially programmed within the bulk by tailoring the ESP profile. These states are also easily reconfigurable over time, providing spatial and temporal freedom that surpasses the limitations of the current valley topological insulator (TI) paradigm, thereby offering greater practicality. Additionally, the mechanism enables a spatially separated topological response, where a localized ESP alters the topological properties in distant regions away from the field. 

To understand this mechanism, we analyze the effect of the ESP on the Jackiw–Rebbi (JR) mass term, revealing how it induces a cross-zero point in real space that generates topological states. Based on this, we construct an extended bulk-edge correspondence to explain both the conventional scenarios and the novel phenomena introduced by our findings. These results are experimentally and numerically verified in mechanical systems, with extensions to electronic and acoustic systems, demonstrating their universal validity. Furthermore, these findings introduce the concept of ``topological lithography,'' enabling the creation of arbitrary, reconfigurable topological patterns for a wide range of systems with external fields.

\subsection*{Model}\label{sec2}

Our system is based on a common hexagonal lattice consisting of two sub-lattices, as shown in Fig. 2(a), where $(i,j)$ is the index of unit cells. In this structure, a spatially varying external potential can be applied and superimposed with the original onsite potential, as illustrated in Fig. 2(b). Within the tight-binding framework, the Hamiltonian is given by:
\begin{equation}\label{eq1}
    H=\sum_{i,j,k}^{} [ t\sum_{i',j',k'}^{}(u_{i',j',k'}^\dagger u_{i,j,k} +  h.c.) + p_k(i,j)u_{i,j,k}^\dagger u_{i,j,k} ]
\end{equation}
where $u$ and $u^\dagger$ are annihilation and creation operators or mode on each node in tight-binding approximation; $t$ is a constant hopping factor or interaction between different nodes; $\sum_{i'j'k'}^{}$ represents the sum over all connections throughout unit cell $(i,j)$ and sub-lattice type $k$ ($k\in\{\alpha,\beta\}$ with $\alpha,\beta$ denoting two sub-lattice types); $h.c.$ is the Hermitian conjugate term; $p_k(i,j)$ is the final potential at each node produced by its coupling to the external potential. Note that the two different sub-lattices exhibit distinct couplings to external field, resulting in distinct final potentials on them, as shown by the two envelope curves in Fig. 2(b) bottom panel.

\begin{figure}[htbp]
\begin{center}
\centerline{\includegraphics[width=1.02\linewidth]{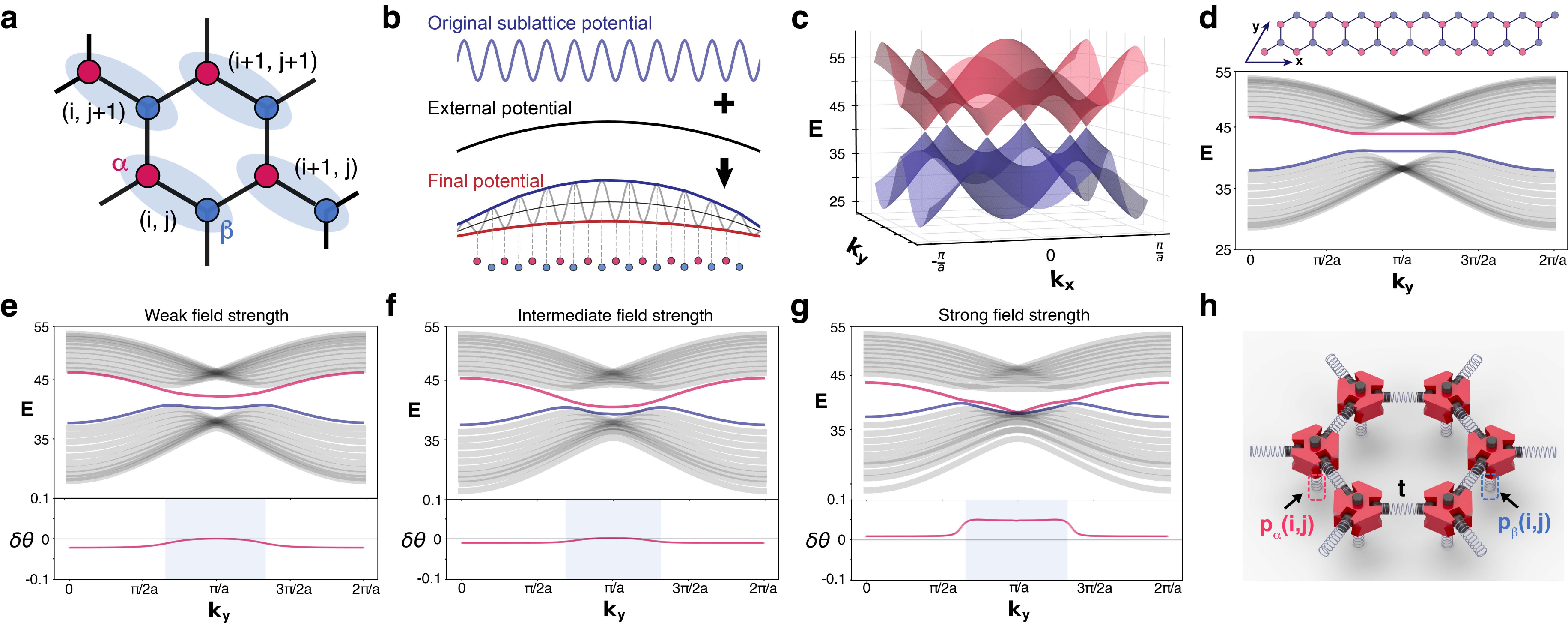}}
\caption{\label{fig.2}(a) Schematics of basic hexagonal lattice. The index $(i,j)$ indicates one unit cell, which contains two sub-lattices, $\alpha$ and $\beta$. (b) Schematics of original lattice potential without external field (top), ESP field (middle) and final potential (bottom). Note that two sublattices have two different couplings to the external field, resulting in two different final potentials indicated by blue and red envelope curves. (c) Band structure of the original hexagonal lattice. Two valleys exist in the 1st BZ, which form the basis of topological protection. (d)  bands of a quasi-1D stripe with zigzag boundaries without the ESP (bottom panel) and the corresponding real-space structure (top panel). The stripe is finite along x-axis and infinite along y-axis. (e)-(g) show the band structure (top) and the eigenvector phase difference (bottom) in the 1st BZ with the increase of ESP strength from low (e) to intermediate (f) and to high (g). The valley regions are marked by the shaded area and a significant increase of $\delta \theta$ appears in (g). (h) Illustration of experimental structure. The hopping parameter $t$ is implemented using identical horizontal springs, while the onsite potential $p(i,j)$ is achieved through vertical grounded springs. The vertical springs can be continuously adjusted in length to realize the desired ESP profiles.}
\vspace{-0.5cm}
\end{center}
\end{figure}

\noindent Solving Eq. (\ref{eq1}) without external potential yields the well-known band structure in Fig. 2(c), featuring two Dirac cones with opposite Berry curvatures in the first Brillouin zone (1st BZ)\cite{schaibleyValleytronics2DMaterials2016}, which forms the foundation of valley protection. We assemble the unit cells into a quasi-1D stripe that is finite along the x-direction and infinite along the y-direction, as shown in the upper panel of Fig. 2(d). Because different sub-lattices $\alpha$ and $\beta$ typically exhibit distinct onsite potentials even without external potential, as demonstrated in Fig. 2(b) top panel, this breaks the degeneracy and opens a gap in the band spectrum, as shown in Fig. 2(d).

\subsection*{ESP Induced Topological States}\label{sec3}

Next, we introduce the external scalar potential (ESP) field, which is analogous to an electrostatic field in electric systems. Adding a uniform external potential merely shifts the entire band structure, which is a trivial effect. To achieve nontrivial behavior, we apply a spatially varying potential field, as illustrated in Fig. 2(b). We begin with a simple quadratic external scalar potential (ESP): $p_k(x)\equiv(p_{\alpha},p_{\beta})(2-\gamma(x-5.5)^2)$, where $\gamma$ determines the ESP strength or the peak height of the external potential, and $x=5.5$ is around the center of the system, ensuring that the potential peak is located near the center. $p_{\alpha}$ or $p_{\beta}$ represents the original onsite potential at each sub-lattice node, which multiplies the external potential to obtain the final potential. For example, $p_{\alpha}$ or $p_{\beta}$ can be associated to the charge of each site, which multiplies an external electrostatic potential to produce the final potential energy. 

As $\gamma$ is gradually increased from zero, the corresponding spectrum gradually evolves from Fig. 2(d) to (e-g). Panels (e)–(g) represent weak, intermediate, and critical values of $\gamma$, respectively. A spectrum curve above the gap, marked in red, is progressively pushed downward as $\gamma$ increases, eventually crossing the gap and connecting to the lower bands as an edge state curve. The bottom panels in Figs. 2(e–g) show the phase difference between small plaquettes in the first Brillouin zone (1st BZ). This phase angle, $\delta\theta$, is derived from the calculation of the Wilson loop:
\begin{equation*}
    \delta \theta(n)=\text{Arg}(\langle u(k_{n,p})|u(k_{n,p+1})\rangle)
\end{equation*}
where $p$ is index of small plaquette in the 1st BZ. Because $\delta\theta$ contributes to the band topology (see SI for details)\cite{alexandradinataWilsonloopCharacterizationInversionsymmetric2014}, an apparent jump in (g) indicates a significant phase variation, suggesting a topological transition at the critical $\gamma$ strength. Next, we aim to quantitatively illustrate this transition.

Due to the time-reversal symmetry of our system, the net topological invariant equals 0. Therefore, an alternative approach is required to characterize the topology. To address this, we introduce a real-space topological marker, referred to as the local topological marker (LTM), inspired by previous real-space topological representations \cite{meierObservationTopologicalAnderson2018,cerjanLocalMarkersCrystalline2024,sykesLocalTopologicalMarkers2021,mondragon-shemTopologicalCriticalityChiralSymmetric2014,linRealspaceRepresentationWinding2021}. The LTM can be expressed by commutation relation of eigenstates mapping in real space:
\begin{equation}\label{eq2}
    \mathcal{C}(x)=\mathop{\rm{average}}\limits_{\alpha,\beta}(Q_{\alpha\beta}[Q_{\beta\alpha},X]_x+Q_{\beta\alpha}[X,Q_{\alpha\beta}]_x)
\end{equation}
where $X$ is position operator; $Q_{\alpha\beta}=\Gamma_\alpha(P_+-P_-)\Gamma_\beta$; $\Gamma_\alpha, \Gamma_\beta$ denote the sub-lattice projectors which project the states onto each sublattice, and $P_\pm$ are the projectors onto the bands where $\pm$ means above/below the gap (see SI for details). Eq. (\ref{eq2}) gives the real-space topological marker of the in-gap mode. With the Jackiw-Rebbi mechanism shown in the next session, we will prove that topologically protected states can be induced by the ESP field. Fig. 2(h) displays the mechanical system we used to experimentally verify our findings. In this setup, inter-node hopping is achieved using identical in-plane springs, while the ESP is implemented through adjustable grounded vertical springs (see SI for additional details). Beyond this mechanical system, we also numerically verify our findings in both electric and acoustic systems.

\begin{figure}[htbp]
\begin{center}
\centerline{\includegraphics[width=1.0\linewidth]{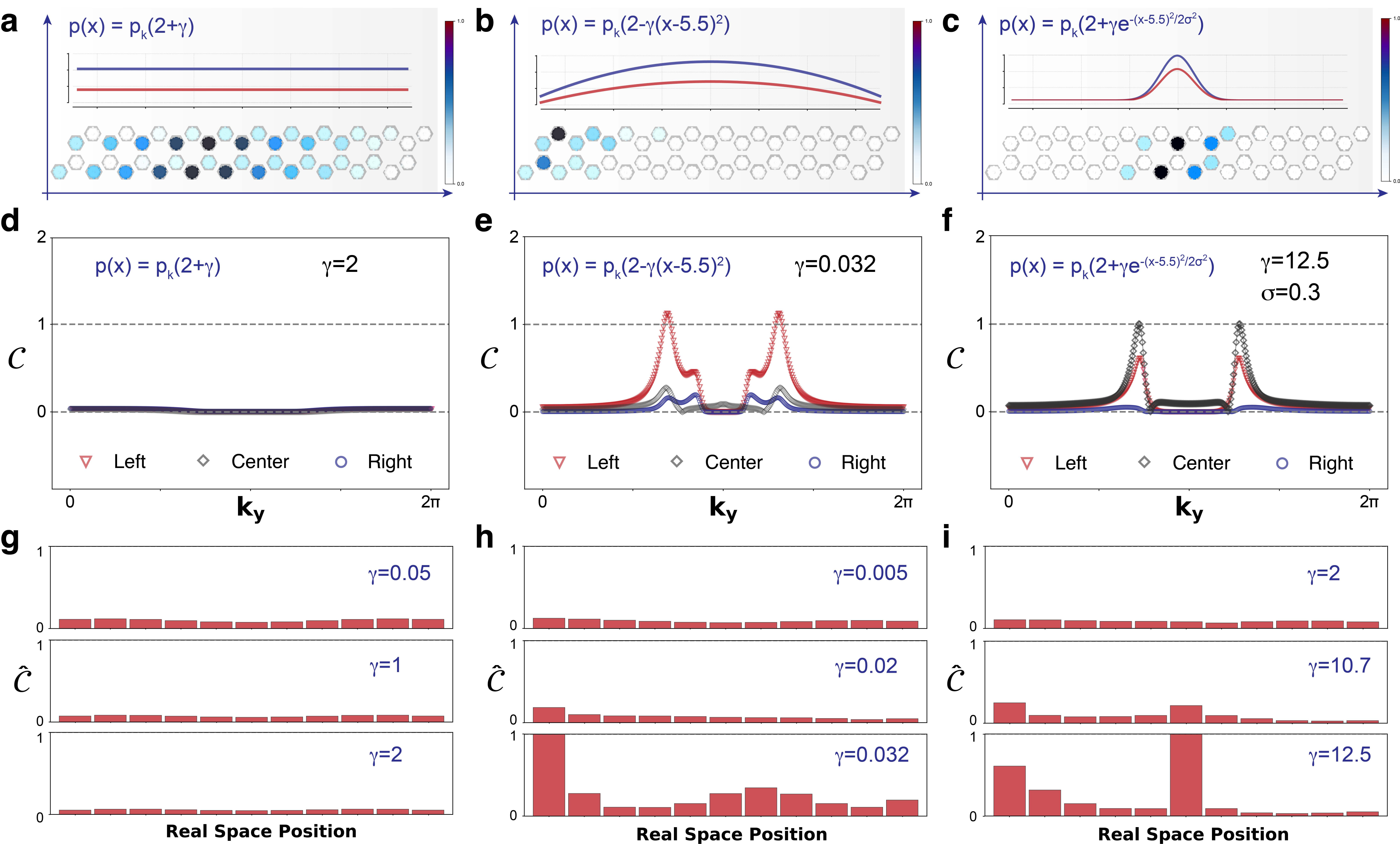}}
\caption{\label{fig.3}(a)–(c) display 1D stripe structures with three different ESP fields (top panels) and their corresponding in-gap edge states (bottom panels). (a) shows a uniform ESP field, (b) a quadratic ESP field, and (c) a Gaussian ESP field. Note that the blue and red curves represent the potentials of two different sublattices, arising from their distinct coupling to the same external field. (d)–(f) present the real-space local topological marker (LTM) $\mathcal{C}$ with respect to $k_y$ at the critical ESP strengths of (a)–(c), respectively. The three curves correspond to three different locations: the left boundary, center, and right boundary. (g)–(i) illustrate the real-space variation of the valley-averaged LTM, $\mathcal{\hat{C}}$, with an increasing ESP strength. The results at the critical strength (bottom panels) match (d)-(f) very well.}
\vspace{-0.5cm}
\end{center}
\end{figure}

We compare several distinct potential profiles in Fig. 3(a)–(c) to illustrate the effects of the ESP field (see more details in SI section III). (a) shows a spatially uniform potential, (b) represents a quadratic profile, and (c) depicts a Gaussian profile, which closely resembles the electric potential typically generated by electric probes in practice. The blue and red potential curves in each panel correspond to the two sublattices, respectively, which exhibit different couplings, $p_\alpha$ and $p_\beta$, to the external potential. Clearly, Fig. 3(a) exhibits a bulk mode, indicating that no topological modes are induced by the uniform potential. In contrast, Figs. 3(b) and 3(c) display localized modes, either at the left edge or the center, which will later be demonstrated as topological modes induced by the non-uniform ESP field.

Fig.3(d)-(f) illustrate the local topological marker (LTM) $\mathcal{C}$ at three locations: the left boundary, the right boundary, and the center. For the uniform ESP shown in Fig. 3(d), $\mathcal{C}=0$ throughout the system, indicating the absence of topological states for this trivial situation. However, for the quadratic profile in Fig. 3(e), $\mathcal{C}$ reaches 1 at the left edge for the critical strength $\gamma=0.032$, signifying the emergence of a topological edge state at the left boundary, while the right boundary remains inactive. This occurs because the ESP field only drives one edge state of the sublattice across the gap (i.e., the red band in Fig. 2(g)), while the other sublattice counterpart (the blue band in Fig.2(g)) is pushed into the bulk bands. As a result, the quadratic ESP can create a topological edge state without altering the bulk symmetry, offering greater design freedom compared to conventional approaches. However, this edge state is still confined to the geometric boundary, similar to the conventional situation, which limits its applicability in broader contexts.

By contrast, and most strikingly, in Fig. 3(f), corresponding to the Gaussian profile, $\mathcal{C}$ reaches 1 at the center, aligning with the potential peak location. This demonstrates that the ESP field can induce a topological state at the peak location rather than at the system edge, highlighting the possibility of freely tunable topological state locations! This phenomenon offers an exciting perspective: Gaussian-like potentials, which are commonly observed in practical settings, allow the creation of topological states at specific locations without reconstructing the lattice or fabricating a structural interface. This could greatly enhance the diversity, tunability, and practical application of topological states and topological materials.

Also, note that the left geometric edge in Fig. 3(f) is also influenced by the ESP, acquiring a nontrivial $\mathcal{C}$. Although it is weaker than in the Gaussian peak region, clear topological protection is still observed, as will be shown later. Since the left edge is far from the Gaussian peak and nearly unaffected by the Gaussian potential, this topological effect demonstrates a remote topological phenomenon: the topologically states of the left geometric edge can be controlled by a spatially separated Gaussian potential. As detailed in the SI, this remote influence can extend up to approximately 60 lattice constants.

Fig. 3(g)-(i) present the valley-averaged $\mathcal{\hat{C}}$ in real space at different $\gamma$ values, corresponding to the structures in Fig. 3(a)-(c), respectively. The results shown in Fig. 3(g)–(i), particularly at the critical $\gamma$ values in bottom panels, align excellently with those in Fig. 3(d)-(f), providing another direct evidence for this field-induced topological effect in real-space.

\subsection*{Jackiw-Rebbi Mechanism and Extended Bulk-Edge Correspondence}

The local topological marker (LTM) $\mathcal{C}$ provides a real-space characterization of the system's topology. The mechanism of such topological states can be understood in terms of a Jackiw-Rebbi (JR) mass term $m_\text{eff}$: the spatially-varying ESP drives $m_\text{eff}$ to cross 0 in the topological region. Near the Dirac cone area of the hexagonal lattice, the Hamiltonian can be written as:
\begin{equation}\label{eq3}
    H_\tau(\bm{r},\bm{k})=v(\tau k_x \sigma_x + k_y \sigma_y)+m_{\text{eff}}(\bm{r},k_y)\sigma_z + \bm{V(\bm{r})}
\end{equation}
where, $\tau$ indicates the valley index $K$ or $K'$ in the hexagonal lattice, and $V(\bm{r})$ is a scalar potential term that only shifts the bands. $m_{\text{eff}}(\bm{r},k_y)$ is a mass term in JR model. In the original lattice without ESP, $m_{\text{eff}}$ is constant. However, in our system it reorganizes into a modified mass $m_{\text{eff}}$ varying with $\bm{r},k_y$ due to the introduction of ESP, Bloch condition along y-axis and the influence of boundary condition (see SI for more details). As we manipulate the ESP field, $m_{\text{eff}}(\bm{r},k_y)$ changes accordingly and may cross 0, thereby generating a Jackiw-Rebbi (JR) domain wall state. Combined with the valley Hall picture, states on the $m_\text{eff}=0$ domain wall are naturally linked to valley protection and an extended bulk-edge correspondence. Specifically, the valley Chern number is related to the sign of $m_\text{eff}$: $C_\tau(\bm{r})=\frac{\tau}{2}\text{sgn}(m_{\text{eff}})$. This relationship provides the JR domain wall state with valley-topological protection: on both sides of the edge, $m_{\text{eff}}$ has opposite signs, giving opposite valley Chern numbers for a given valley and thereby supporting a valley-protected interface state, which is the basic mechanism of a Valley Hall Insulator (VHI) interface. The effect of the ESP field on the position of topological states—via its influence on $m_{\text{eff}}$—aligns with the LTM characterization in Fig. 3 and explains its underlying mechanism (see SI for additional details).

Moreover, this framework offers an extended bulk-edge correspondence in valley materials:
 
\begin{equation}\label{eq4}
    n_{\text{valley}}=\frac{1}{2}|\text{sgn}(m_{\text{eff},\text{side\ 1}})-\text{sgn}(m_{\text{eff},\text{side\ 2}})|
\end{equation}
where side 1 and 2 refer to the two sides of a chosen reference interface, $n_{\text{valley}}$ represents the number of allowed valley topological states between side 1 and 2, and $\text{sgn}(m_{\text{eff},1/2})=\pm 1$ indicates the sign of the effective JR mass term on each side. The relation in Eq.(\ref{eq4}) places conventional Valley Hall interfaces and the ESP-written domain walls within the same real-space mass-sign criterion. 

\subsection*{Quadratic ESP Inducing Topological Protection at Geometric Edge}\label{sec4}

In this section, we experimentally verify the nontrivial edge states induced by a quadratic ESP field at a geometric boundary, as shown in Fig. 3(b), (e), and (h). To achieve this, we construct a mechanical vibrational system based on Fig. 2(h) that is described by Eq.(\ref{eq1}).

The experimental system is illustrated in Fig. 4(a), where a quadratic ESP field is distributed along the x-axis while remaining uniform along the y-axis. The ESP profile is designed to match the one shown in Fig.3(b), with $p_{\alpha}=42.67N/m$ and $p_{\beta}=32N/m$ respectively. For each node, we measure only its vertical out-of-plane vibration using a laser displacement sensor with an accuracy of $10\mu m$. In the horizontal plane, the lattice nodes are connected by identical stretched springs, which serve as the hopping parameter $t$ (see SI for more details). The ESP at each lattice site is implemented using a grounded, length-controllable vertical spring. By controlling the lengths of these springs, an overall quadratic ESP is realized.

\begin{figure}[htbp]
\begin{center}
\centerline{\includegraphics[width=0.95\linewidth]{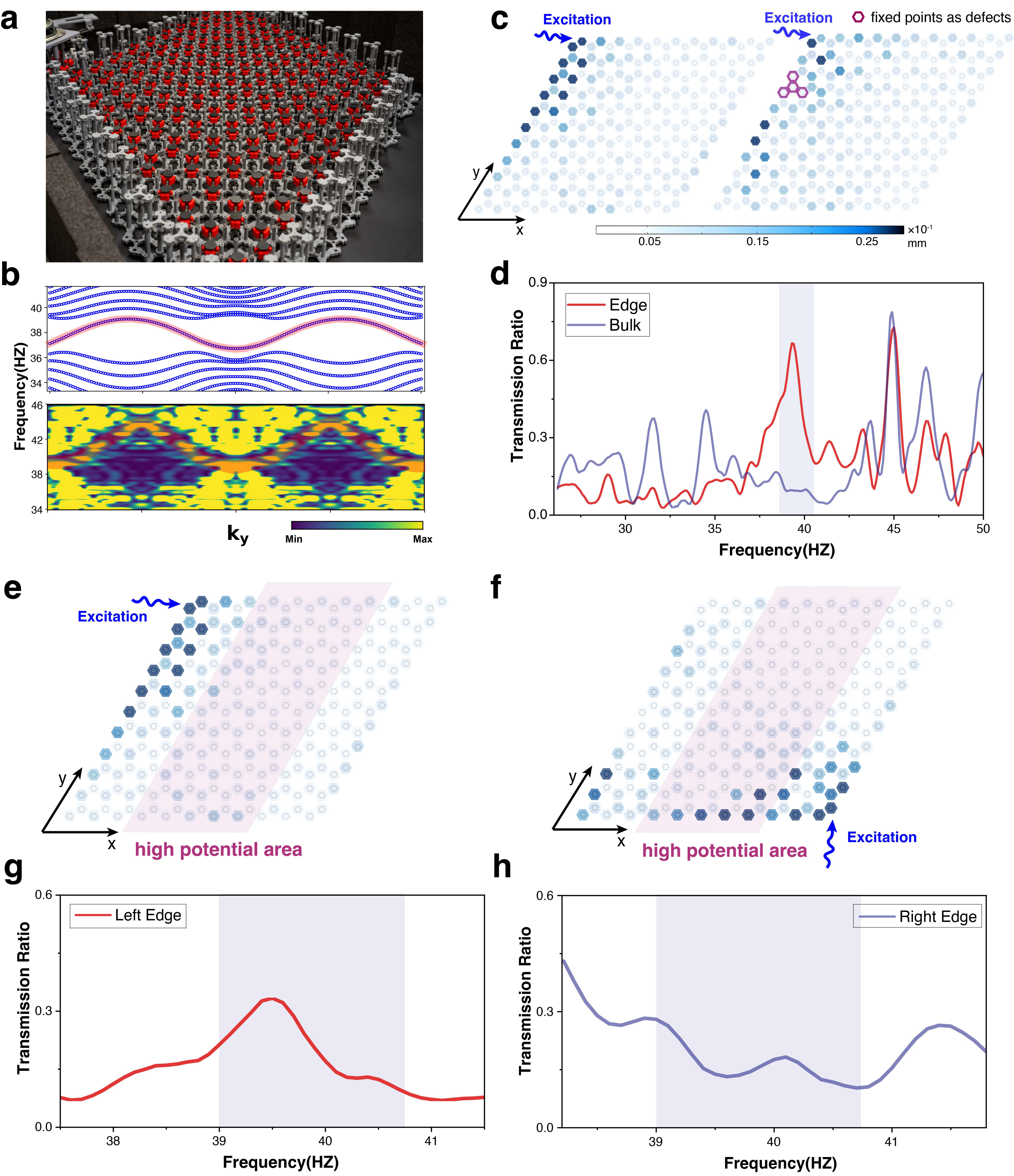}}
\caption{\label{fig.4}Topological states on geometric boundaries induced by quadratic ESP. (a) The experimental setup, where the vertical vibrations are measured by laser sensors with the accuracy of $10\mu m$. (b) The simulated (top) and measured (bottom) band spectrum, with topological edge states highlighted by light-red color. They agree very well. (c) Vibrational modes of the topological edge states without defects (left) and the edge state with defects introduced by fixed points (right). The topological protection is clearly evident. (d) The transmission ratio of edge and bulk modes measured from (c), using sensors placed at the left boundary and within the bulk. The band gap is indicated by the shaded area, and a clear transmission peak is observed at the left boundary. (e, f) Display vibrational modes from two different excitation points, respectively. Clearly, the left boundary exhibits topological protection, while the right boundary does not. The shaded region represents areas with high potential values in the quadratic ESP. (g, h) Show the corresponding transmission ratios from (e, f), with the band gap highlighted by the shaded area.}
\vspace{-0.5cm}
\end{center}
\end{figure}

For the edge states induced by the quadratic ESP, we simulated and measured the dispersion spectrum in Fig. 4(b). The top panel presents the simulation, while the bottom panel displays the experimental results. The nontrivial edge-state band is highlighted in red, and the experimental measurements exhibit excellent agreement with the simulations. The band gap is located around 39.9 Hz. Based on this, the corresponding topological modes were measured, as shown in Fig. 4(c), where the left panel depicts edge modes without defects and the right panel shows edge modes with boundary defects by fixed nodes. It can be observed that the edge state does not scatter into the bulk, even when boundary defects are introduced, demonstrating robust topological protection. Fig. 4(d) presents the bulk and edge transmission measured from a boundary and a bulk probe respectively. The band gap region is highlighted by the shaded area, wherein the bulk probe exhibits a very low transmission ratio, while the edge probe reveals a strong transmission peak—a typical feature of topological edge states.


Our experiments further confirm the predicted asymmetry of the topological edge states, with the results summarized in Fig. 4(e-h). Exciting the left corner launches a robust edge mode that propagates primarily along the nontrivial geometric edge (4e, left edge). In stark contrast, the right boundary does not support such a mode within the band gap (4f, right edge). This directly demonstrates that the topological protection on boundary is related to specific sublattice, similar to the boundary phenomena in VHI\cite{zhao_elastic_2022,liu_tunable_2018}. The transmission spectra (g,h) corroborate this, showing a clear transmission peak for the left edge within the band gap, a feature completely absent on the right edge. These results provide strong experimental validation of our model.

\subsection*{Gaussian ESP Inducing Topological States Within the Bulk and Its Remote Effect}

Next, we adjust the ESP field into a Gaussian profile with its peak located at the center of the system. The experimental system is similar to that of the quadratic case, with a Gaussian ESP field applied along the x-direction while remaining uniform along the y-direction. The band spectrum is again simulated and measured, with the results presented in Fig. 5(a). The top and bottom panels show the simulation and experimental results, respectively, with the band of topological state highlighted in red. The agreement between simulation and experiment is excellent.

Fig. 5(b) offers a direct comparison of the topological state measured within the gap, contrasting the simulated mode at 34 Hz (left) with the experimental measurement at 31.06 Hz (right). The remarkable agreement between the predicted and observed modes provides strong experimental evidence for the existence of the topological ``edge'' state located at the center, rather than at the geometric edge.

\begin{figure}[htbp]
\begin{center}
\centerline{\includegraphics[width=1.1\linewidth]{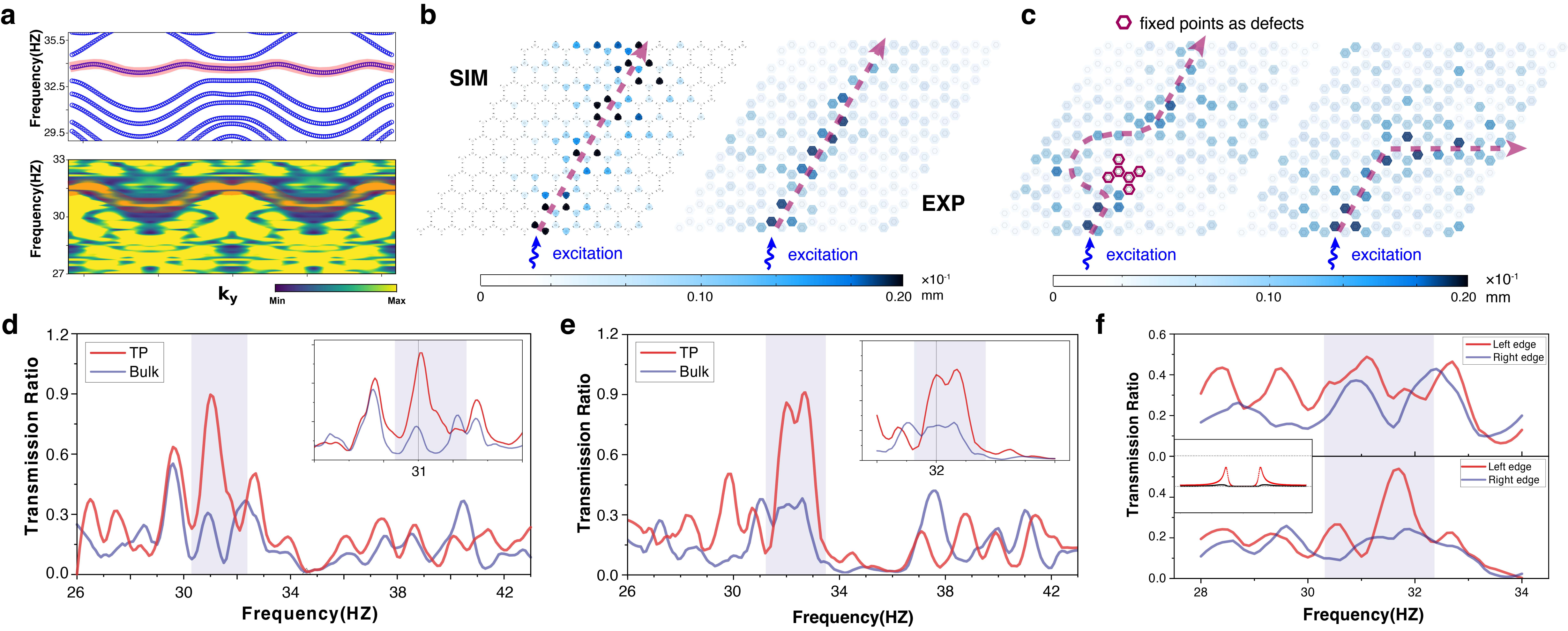}}
\caption{\label{fig.5}Topological states inside the bulk induced by Gaussian ESP. (a) Band spectrum obtained from simulation (top) and experiment (bottom), with the topological states highlighted in light red. The simulation and experimental results show excellent agreement. (b) Vibrational modes of the topological states obtained from simulation (left) and experiment (right). The topologically protected pathway is indicated by the dashed arrow. (c) Experimentally measured modes of the topological states in the presence of defects (left) and designed as an L-shaped pattern (right). The propagation direction is again indicated by the dashed arrow, demonstrating topological protection clearly. (d, e) Transmission ratios measured by a sensor located inside the topologically protected pathway (labeled as ``TP'') and another sensor outside the pathway but still inside the bulk (labeled as ``bulk''), respectively. The band gap is highlighted with a shaded area and the insets show zoomed-in pictures. (f) Transmission ratios measured at the left and right geometric edges without (top) and with defects (bottom). The band gap is marked by the shaded area. In the top panel, topological protection on the left edge is absent, but it becomes clear after defects are introduced in the bottom panel. The inset shows the LTM spectrum of left and right boundaries under critical $\gamma$ (taken from Fig. 3(f)), revealing topological protection at the left edge.}
\vspace{-0.5cm}
\end{center}
\end{figure}


We once again confirm the robust topological protection of the topological states against both structural defects and geometric paths. As shown in Fig. 5(c), the state propagates unimpeded around a series of fixed-node defects (left) and seamlessly navigates a turning corner designed into the ESP field (right). In both cases, the measured mode profiles demonstrate the robustness of the topological state, propagating smoothly around the defects and along the turning corner. This robustness is further quantified by the transmission spectra in Fig. 5(d) and 5(e), corresponding to the defective and turning-path configurations, respectively. The band gap region is highlighted by shaded area, with insets providing a magnified view. In both scenarios, the ``TP'' probe located within the path of the topological state records a high-transmission peak within the band gap, confirming robust topological protection (i.e., TP). Conversely, the ``bulk'' probe located at a normal bulk location but outside the path measures strongly suppressed transmission, emphasizing a sharp contrast to the topologically protected path.

These results provide compelling experimental evidence of valley-protected ESP-induced topological states, in excellent agreement with theoretical predictions in Fig. 3(c). Most strikingly, the topological states are located at the Gaussian peak area and do not need to coincide with geometric edges. Owing to the easily tunable nature of the external potential field, this feature offers substantial flexibility for designing topological transport channels with desirable patterns—a concept we refer to as ``topological lithography''. Furthermore, the external field can be varied over time, providing additional temporal freedom.

Notably, our simulations in Fig. 3(i) suggest that the central Gaussian ESP imparts a weak topological state to the left geometric boundary, an effect absent on the right geometric boundary. To experimentally verify this asymmetry, we measure the transmission spectra for both boundaries, as shown in Fig. 5(f). Without defects (top panel), both boundaries exhibit similarly low transmission within the band gap, with no discernible conducting channel, leaving the topological difference unresolved. However, this hidden topological difference becomes evident when defects or fixed boundary nodes are introduced into the left and right boundaries, respectively, and the transmission spectra are re-measured. The results (bottom panel) reveal a striking contrast: the left boundary sustains a distinct transmission peak within the gap, whereas the right boundary’s transmission remains fully suppressed. This defect resilience on the left geometric boundary, absent on the right, provides clear experimental evidence for the remote topological influence produced by the Gaussian ESP within the bulk. Also note that this remote effect does not extend infinitely and becomes negligible when the distance exceeds about 60 lattice constants (see SI Section IV for more details).

\subsection*{General Extension to Various Systems}

In the work presented above, we have investigated the influence of two types of ESP on the system's topology. Although all validations were conducted in a mechanical network, the same mechanism is general and can, in principle, be implemented in other wave systems. For example, we extend our system to an electronic system, such as a graphene-like hexagonal lattice (e.g., h-BN, h-BP, GaN, etc.), as well as to an acoustic system. In Fig. 6(a), we present a schematic diagram illustrating a potential realization of such novel topological states in a graphene-like structure via an ESP. The onsite potential in the mechanical network can be replaced by an electric field that varies spatially along the x-axis. In this scenario, $p_{\alpha}$ and $p_\beta$ correspond to the different couplings of sub-lattice atoms, which are multiplied by the external electric potential to determine the final potentials. The external potential could be achieved by designing specific structures of electrodes or the substrate medium\cite{wang2024locally,forsythe2018band}, as shown in (a).

\begin{figure}[htbp]
\begin{center}
\centerline{\includegraphics[width=1.1\linewidth]{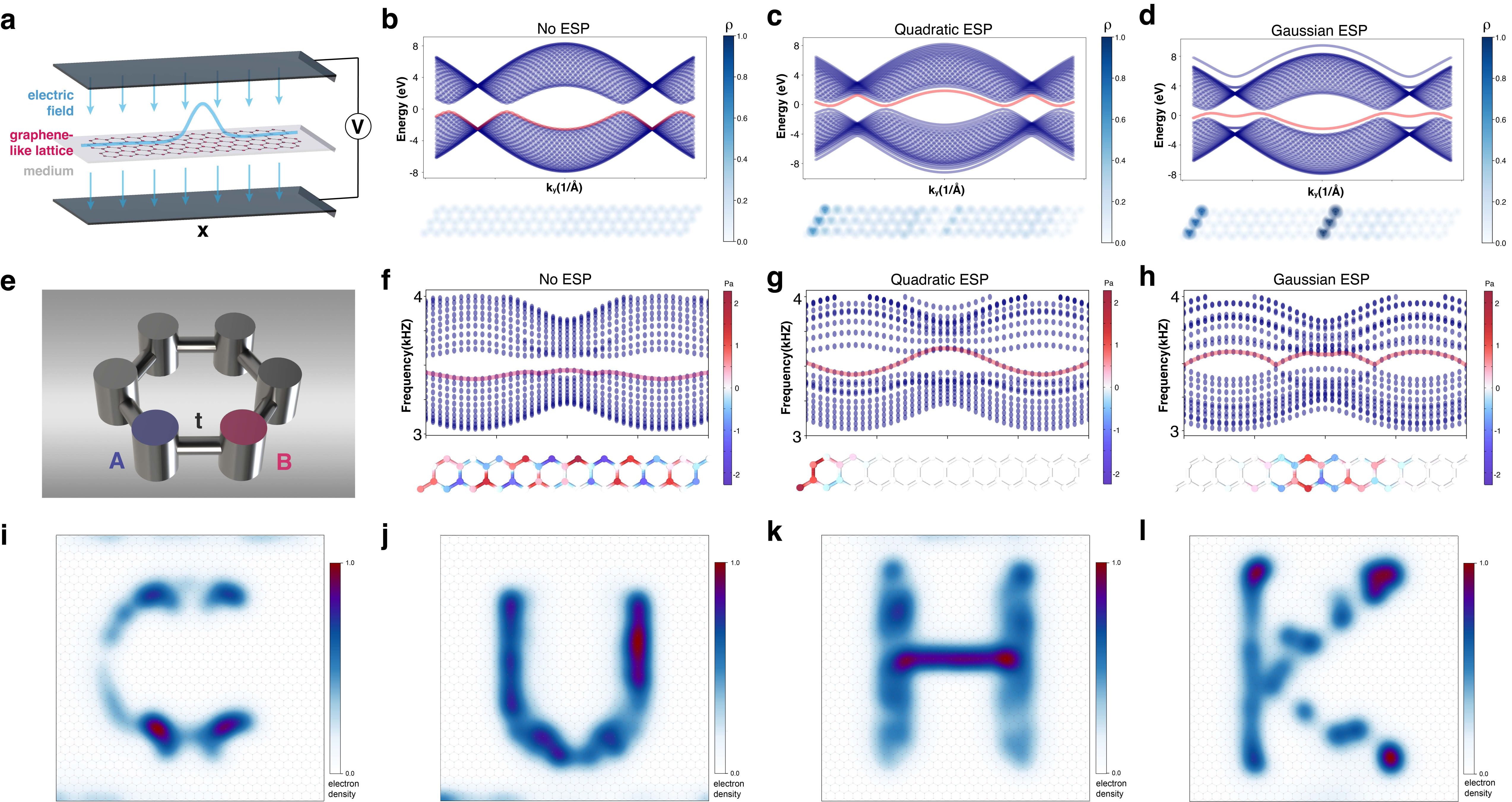}}
\caption{\label{fig.6}Inducing topological states in graphene-like electronic systems (a–d) and acoustic systems (e–h) through an ESP field. (a) A schematic illustration of the realization in an electronic system. (b–d) Show simulations of the electron band structure without ESP (b), with a quadratic-shaped ESP field (c), and with a Gaussian-shaped ESP field (d). The bottom panels display the real-space modes of the topological states. (e) A schematic of one hexagonal cell of an acoustic lattice composed of resonant cavities. (f–h) Present simulations of the acoustic band structure without ESP (f), with a quadratic-shaped ESP field (g), and with a Gaussian-shaped ESP field (h). The bottom panels exhibit the real-space modes of the topological states. Results from both electronic and acoustic systems align well with those from the previously studied mechanical system. (i–l) Display simulations of the lithography effect of topological modes forming four letter-shaped patterns (`C', `U', `H', `K') in the same electronic system under corresponding Gaussian ESP fields.}
\vspace{-0.5cm}
\end{center}
\end{figure}

We use the tight-binding model to calculate the electronic eigenmode spectrum on this graphene-like structure, with the results shown in Fig. 6(b–d). Specifically, Fig. 6(b) illustrates the situation without an ESP field. The top panel shows the band spectrum, with the band of interest highlighted in red. The bottom panel shows a real-space mode corresponding to the red band, where no topological state is observed. However, upon applying quadratic and Gaussian ESP fields, as shown in Figures 6(c) and 6(d), the red band closes the gap at the critical field strength $\gamma$, leading to a topological state at either the left geometric boundary (Fig. 6(c)) or at both the center and left boundary (Fig. 6(d)). These results are in excellent agreement with those from the previous mechanical system, demonstrating the validity of the mechanism in an electronic system (see the Methods section and SI for more details).

Similar results are also confirmed in an acoustic lattice, where Fig. 6(e) illustrates a hexagonal basic  block composed of acoustic resonant cavities. The sub-lattice difference is implemented by using two different types of resonant cavities, A and B, depicted in different colors. Figures 6(f)–(h) present the results for zero, quadratic, and Gaussian ESP fields, respectively. The topological bands are again highlighted in red, and the corresponding modes are displayed in the bottom panels. These topological modes closely replicate the behavior observed in the mechanical counterparts, demonstrating the general validity of the principle once again.

Note that the spectrum in Fig. 6 may appear upside-down compared to the mechanical system in Fig. 2. This difference arises from the eigenvalue of $-\omega^2$ in mechanical systems, which carries a negative sign. However, this does not affect any of the topological mode properties (see SI for additional discussion and extensions).

To showcase the potential of our topological lithography approach, Fig. 6(i)–(l) demonstrates four different letter patterns created within the same graphene-like electronic system, induced by four distinct Gaussian ESP electric fields. The letters ``CUHK'' represent the abbreviation for our university, The Chinese University of Hong Kong. This vividly demonstrates the ability to create arbitrary topological-state pathways at arbitrary locations, surpassing the limitations of geometric boundaries, through our topological lithography technique. Consequently, our work not only deepens the theoretical understanding of topological states but also significantly enhances the diversity, tunability, and practical implementation of topological states and materials.

\subsection*{Discussion and Conclusion}\label{sec5}

In this work, we introduce a novel mechanism to manipulate valley topology in hexagonal lattice systems using spatially varying external potential fields (ESPs). Unlike conventional valley-topological implementations in which transport channels are structurally defined by interfaces between distinct domains, our approach uses a tunable external field to create and position the effective topological interface within a fixed lattice. Through both quadratic and Gaussian ESPs, we observe topological states at boundaries and within the bulk, with experimental validations in mechanical lattices. The Gaussian ESP, in particular, enables topological states both at its peak location and at distant boundaries, introducing new strategies for topological design beyond the traditional paradigm.

Our results also demonstrate the concept of ``topological lithography,'' enabling the creation of arbitrary, reconfigurable patterns of topological states, as exemplified by letter-shaped patterns in graphene-like lattices. The universality of this mechanism was experimentally validated in mechanical, and numerically in electronic and acoustic systems, confirming its adaptability across different platforms. These ESP-induced topological states pave the way for designing flexible, scalable devices for applications in wave manipulation, signal routing, and energy transport. To some extent, our discovery is analogous to the field-effect transistor, except instead of controlling charge transport, it manipulates the topological state, which includes but is not limited to charge transport. Therefore, our study not only broadens the understanding of valley physics and topological materials but also establishes a framework for reconfigurable and programmable topological devices. Future work could explore more complex ESP designs and dynamic modulations, unlocking new possibilities for applications in wave physics and multifunctional material systems.

\backmatter

\subsection*{Methods}\label{sec11}

\bmhead{Simulation and experimental structure}

The theoretical calculations presented in this paper were primarily conducted using the numpy module in Python. The experimental simulations were carried out using the Solid Mechanics module in COMSOL Multiphysics.

The main experimental structure is fabricated using a fused-deposition-modeling 3D printer (Raise3D Pro3 Ultra) and a stereolithography 3D printer (ELEGOO Saturn 3 Ultra). The lattice nodes are made from Raise3D PLA[1.75mm] and DD112 light-curving resin. The mass differences introduced by the fabrication methods are negligible (less than 0.2\%). In the experiments, we used PLA to construct the main structure of the lattice, while photo-curable resin, known for its high precision but more complex manufacturing process, was utilized for the spring connectors. In the experiments, the stiffness of the springs was regulated by precisely controlling the length of the threaded connections on the connectors, which in turn adjusted the effective length of the springs. 

\bmhead{Experimental measurements}

The vibration of each node is measured with a CMOS displacement laser sensor GC05-30NW, which has a response time of 1.5ms with repeatability precision 10$\rm{\mu s}$. Combined with a M58 DAQ card, this setup is suitable for measuring small-amplitude vibrations under 200HZ. The experimental oscillation frequencies lie in the range of approximately 25–50 Hz, which is suitable for the requirements of our measuring devices. 

\bmhead{Acknowledgements}
L.X. acknowledges the financial support from GRF-14307422, GRF-14306923, The Chinese University of Hong Kong (CUHK) direct grant 4053582, X. S. acknowledges the financial support from Guangdong Basic and Applied Basic Research Foundation
(Project Nos. 2025B1515020077, 2024A1515030139), NSFC, under the Grant No. T2550093. Correspondence should be addressed to: shenxy66@sysu.edu.cn and xuleixu@cuhk.edu.hk.
\bmhead{Competing interests}
H.N. and L.X. are coauthors of a filed US patent No. 64/143,095, which describes the method used here in.

\bibliography{reference}

@article{agarwalaTopologicalInsulatorsAmorphous2017,
  title = {Topological {{Insulators}} in {{Amorphous Systems}}},
  author = {Agarwala, Adhip and Shenoy, Vijay B.},
  year = {2017},
  month = jun,
  journal = {Phys. Rev. Lett.},
  volume = {118},
  number = {23},
  pages = {236402},
}

@article{alexandradinataWilsonloopCharacterizationInversionsymmetric2014,
  title = {Wilson-Loop Characterization of Inversion-Symmetric Topological Insulators},
  author = {Alexandradinata, A. and Dai, Xi and Bernevig, B. Andrei},
  year = {2014},
  month = apr,
  journal = {Phys. Rev. B},
  volume = {89},
  number = {15},
  pages = {155114},
}

@article{alezziTopologicalFlatBands2024,
  title = {Topological {{Flat Bands}} in {{Graphene Super-Moir}}{\textbackslash}'e {{Lattices}}},
  author = {Al Ezzi, Mohammed M. and Hu, Junxiong and Ariando, Ariando and Guinea, Francisco and Adam, Shaffique},
  year = {2024},
  month = mar,
  journal = {Phys. Rev. Lett.},
  volume = {132},
  number = {12},
  pages = {126401},
}

@article{ashidaNonHermitianPhysics2020,
  title = {Non-{{Hermitian Physics}}},
  author = {Ashida, Yuto and Gong, Zongping and Ueda, Masahito},
  year = {2020},
  month = jul,
  journal = {Advances in Physics},
  volume = {69},
  number = {3},
  eprint = {2006.01837},
  primaryclass = {cond-mat, physics:quant-ph},
  pages = {249--435},
}

@article{barikTopologicalQuantumOptics2018,
  title = {A Topological Quantum Optics Interface},
  author = {Barik, Sabyasachi and Karasahin, Aziz and Flower, Christopher and Cai, Tao and Miyake, Hirokazu and DeGottardi, Wade and Hafezi, Mohammad and Waks, Edo},
  year = {2018},
  month = feb,
  journal = {Science},
  volume = {359},
  number = {6376},
  pages = {666--668},
}

@book{bernevigTopologicalInsulatorsTopological2013,
  title = {Topological {{Insulators}} and {{Topological Superconductors}}},
  author = {Bernevig, B. Andrei},
  year = {2013},
  month = dec,
}

@article{biancoMappingTopologicalOrder2011a,
  title = {Mapping Topological Order in Coordinate Space},
  author = {Bianco, Raffaello and Resta, Raffaele},
  year = {2011},
  month = dec,
  journal = {Phys. Rev. B},
  volume = {84},
  number = {24},
  pages = {241106},
}

@article{brendelPseudomagneticFieldsSound2017,
  title = {Pseudomagnetic Fields for Sound at the Nanoscale},
  author = {Brendel, Christian and Peano, Vittorio and Painter, Oskar J. and Marquardt, Florian},
  year = {2017},
  month = apr,
  journal = {Proceedings of the National Academy of Sciences},
  volume = {114},
  number = {17},
  pages = {E3390-E3395},
}

@article{caceres-aravenaCompactTopologicalEdge2024,
  title = {Compact {{Topological Edge States}} in {{Flux-Dressed Graphenelike Photonic Lattices}}},
  author = {{C{\'a}ceres-Aravena}, Gabriel and Nedi{\'c}, Milica and Vildoso, Paloma and Gligori{\'c}, Goran and Petrovic, Jovana and Maluckov, Aleksandra and Vicencio, Rodrigo A.},
  year = {2024},
  month = sep,
  journal = {Phys. Rev. Lett.},
  volume = {133},
  number = {11},
  pages = {116304},
}

@article{cerjanLocalMarkersCrystalline2024,
  title = {Local {{Markers}} for {{Crystalline Topology}}},
  author = {Cerjan, Alexander and Loring, Terry A. and {Schulz-Baldes}, Hermann},
  year = {2024},
  month = feb,
  journal = {Phys. Rev. Lett.},
  volume = {132},
  number = {7},
  pages = {073803},
}

@article{chaudronElectricfieldinducedMultiferroicTopological2024,
  title = {Electric-Field-Induced Multiferroic Topological Solitons},
  author = {Chaudron, Arthur and Li, Zixin and Finco, Aurore and Marton, Pavel and Dufour, Pauline and Abdelsamie, Amr and Fischer, Johanna and Collin, Sophie and Dkhil, Brahim and Hlinka, Jirka and Jacques, Vincent and Chauleau, Jean-Yves and Viret, Michel and Bouzehouane, Karim and Fusil, St{\'e}phane and Garcia, Vincent},
  year = {2024},
  month = may,
  journal = {Nat. Mater.},
  pages = {1--7},
}

@article{chenAcousticWeylPoints2018,
  title = {Acoustic {{Weyl}} Points in a Square Lattice},
  author = {Chen, Ting-Gui and Jiao, Jun-Rui and Dai, Hong-Qing and Yu, De-Jie},
  year = {2018},
  month = dec,
  journal = {Phys. Rev. B},
  volume = {98},
  number = {21},
  pages = {214110},
}

@article{chenElasticQuantumSpin2018,
  title = {Elastic Quantum Spin {{Hall}} Effect in Kagome Lattices},
  author = {Chen, H. and Nassar, H. and Norris, A. N. and Hu, G. K. and Huang, G. L.},
  year = {2018},
  month = sep,
  journal = {Phys. Rev. B},
  volume = {98},
  number = {9},
  pages = {094302},
}

@article{chenVariousTopologicalPhases2023a,
  title = {Various Topological Phases and Their Abnormal Effects of Topological Acoustic Metamaterials},
  author = {Chen, Yan-Feng and Chen, Ze-Guo and Ge, Hao and He, Cheng and Li, Xin and Lu, Ming-Hui and Sun, Xiao-Chen and Yu, Si-Yuan and Zhang, Xiujuan},
  year = {2023},
  journal = {Interdisciplinary Materials},
  volume = {2},
  number = {2},
  pages = {179--230},
}

@article{chiuClassificationTopologicalQuantum2016,
  title = {Classification of Topological Quantum Matter with Symmetries},
  author = {Chiu, Ching-Kai and Teo, Jeffrey C. Y. and Schnyder, Andreas P. and Ryu, Shinsei},
  year = {2016},
  month = aug,
  journal = {Rev. Mod. Phys.},
  volume = {88},
  number = {3},
  pages = {035005},
}

@article{coulaisStaticNonreciprocityMechanical2017,
  title = {Static Non-Reciprocity in Mechanical Metamaterials},
  author = {Coulais, Corentin and Sounas, Dimitrios and Al{\`u}, Andrea},
  year = {2017},
  month = feb,
  journal = {Nature},
  volume = {542},
  number = {7642},
  pages = {461--464},
}

@article{cuiOnChipElasticWave2024,
  title = {On-{{Chip Elastic Wave Manipulations Based}} on {{Synthetic Dimension}}},
  author = {Cui, Zhenxing and Wu, Chaohua and Wei, Qiang and Yan, Mou and Chen, Gang},
  year = {2024},
  month = dec,
  journal = {Phys. Rev. Lett.},
  volume = {133},
  number = {25},
  pages = {256602},
}

@article{dongFlatBandsMagicAngle2021,
  title = {Flat {{Bands}} in {{Magic-Angle Bilayer Photonic Crystals}} at {{Small Twists}}},
  author = {Dong, Kaichen and Zhang, Tiancheng and Li, Jiachen and Wang, Qingjun and Yang, Fuyi and Rho, Yoonsoo and Wang, Danqing and Grigoropoulos, Costas P. and Wu, Junqiao and Yao, Jie},
  year = {2021},
  month = jun,
  journal = {Phys. Rev. Lett.},
  volume = {126},
  number = {22},
  pages = {223601},
}

@article{fanElasticHigherOrderTopological2019,
  title = {Elastic {{Higher-Order Topological Insulator}} with {{Topologically Protected Corner States}}},
  author = {Fan, Haiyan and Xia, Baizhan and Tong, Liang and Zheng, Shengjie and Yu, Dejie},
  year = {2019},
  month = may,
  journal = {Phys. Rev. Lett.},
  volume = {122},
  number = {20},
  pages = {204301},
}

@article{fangRealizingEffectiveMagnetic2012,
  title = {Realizing Effective Magnetic Field for Photons by Controlling the Phase of Dynamic Modulation},
  author = {Fang, Kejie and Yu, Zongfu and Fan, Shanhui},
  year = {2012},
  month = nov,
  journal = {Nature Photon},
  volume = {6},
  number = {11},
  pages = {782--787},
}

@article{hafeziImagingTopologicalEdge2013,
  title = {Imaging Topological Edge States in Silicon Photonics},
  author = {Hafezi, M. and Mittal, S. and Fan, J. and Migdall, A. and Taylor, J. M.},
  year = {2013},
  month = dec,
  journal = {Nature Photon},
  volume = {7},
  number = {12},
  pages = {1001--1005},
}

@article{hastingsTopologicalInsulatorsCalgebras2011,
  title = {Topological Insulators and {{C}}{$\ast$}-Algebras: {{Theory}} and Numerical Practice},
  shorttitle = {Topological Insulators and {{C}}{$\ast$}-Algebras},
  author = {Hastings, Matthew B. and Loring, Terry A.},
  year = {2011},
  month = jul,
  journal = {Annals of Physics},
  series = {July 2011 {{Special Issue}}},
  volume = {326},
  number = {7},
  pages = {1699--1759},
}

@article{huberTopologicalMechanics2016,
  title = {Topological Mechanics},
  author = {Huber, Sebastian D.},
  year = {2016},
  month = jul,
  journal = {Nature Phys},
  volume = {12},
  number = {7},
  pages = {621--623},
}

@article{kaneTopologicalBoundaryModes2014a,
  title = {Topological Boundary Modes in Isostatic Lattices},
  author = {Kane, C. L. and Lubensky, T. C.},
  year = {2014},
  month = jan,
  journal = {Nature Phys},
  volume = {10},
  number = {1},
  pages = {39--45},
}

@article{laiTopologicalPhononicFiber2024,
  title = {Topological {{Phononic Fiber}} of {{Second Spin-Chern Number}}},
  author = {Lai, Hua-Shan and Gou, Xiao-Hui and He, Cheng and Chen, Yan-Feng},
  year = {2024},
  month = nov,
  journal = {Phys. Rev. Lett.},
  volume = {133},
  number = {22},
  pages = {226602},
}

@article{liLocalizedDelocalizedTopological2024,
  title = {Localized and Delocalized Topological Modes of Heat},
  author = {Li, Jiaxin and Xu, Chengxin and Xu, Zifu and Xu, Guoqiang and Yang, Shuihua and Liu, Kaipeng and Chen, Jianfeng and Li, Tianlong and Qiu, Cheng-Wei},
  year = {2024},
  month = aug,
  journal = {Proceedings of the National Academy of Sciences},
  volume = {121},
  number = {35},
  pages = {e2408843121},
}

@article{linObservationTopologicalTransition2024,
  title = {Observation of {{Topological Transition}} in {{Floquet Non-Hermitian Skin Effects}} in {{Silicon Photonics}}},
  author = {Lin, Zhiyuan and Song, Wange and Wang, Li-Wei and Xin, Haoran and Sun, Jiacheng and Wu, Shengjie and Huang, Chunyu and Zhu, Shining and Jiang, Jian-Hua and Li, Tao},
  year = {2024},
  month = aug,
  journal = {Phys. Rev. Lett.},
  volume = {133},
  number = {7},
  pages = {073803},
}

@article{linRealspaceRepresentationWinding2021,
  title = {Real-Space Representation of the Winding Number for a One-Dimensional Chiral-Symmetric Topological Insulator},
  author = {Lin, Ling and Ke, Yongguan and Lee, Chaohong},
  year = {2021},
  month = jun,
  journal = {Phys. Rev. B},
  volume = {103},
  number = {22},
  pages = {224208},
}

@article{liuTunableAcousticValleyHall2018,
  title = {Tunable {{Acoustic Valley--Hall Edge States}} in {{Reconfigurable Phononic Elastic Waveguides}}},
  author = {Liu, Ting-Wei and Semperlotti, Fabio},
  year = {2018},
  month = jan,
  journal = {Phys. Rev. Appl.},
  volume = {9},
  number = {1},
  pages = {014001},
}

@article{loTopologyNonlinearMechanical2021,
  title = {Topology in {{Nonlinear Mechanical Systems}}},
  author = {Lo, Po-Wei and Santangelo, Christian D. and Chen, Bryan Gin-ge and Jian, Chao-Ming and Roychowdhury, Krishanu and Lawler, Michael J.},
  year = {2021},
  month = aug,
  journal = {Phys. Rev. Lett.},
  volume = {127},
  number = {7},
  pages = {076802},
}

@article{maExperimentalDemonstrationDualBand2021a,
  title = {Experimental {{Demonstration}} of {{Dual}}-{{Band Nano}}-{{Electromechanical Valley}}-{{Hall Topological Metamaterials}}},
  author = {Ma, Jingwen and Xi, Xiang and Sun, Xiankai},
  year = {2021},
  month = mar,
  journal = {Adv. Mater.},
  volume = {33},
  number = {10},
  pages = {2006521},
}

@article{maNonlinearTopologicalMechanics2023,
  title = {Nonlinear {{Topological Mechanics}} in {{Elliptically Geared Isostatic Metamaterials}}},
  author = {Ma, Fangyuan and Tang, Zheng and Shi, Xiaotian and Wu, Ying and Yang, Jinkyu and Zhou, Di and Yao, Yugui and Li, Feng},
  year = {2023},
  month = jul,
  journal = {Phys. Rev. Lett.},
  volume = {131},
  number = {4},
  pages = {046101},
}

@article{meierObservationTopologicalAnderson2018,
  title = {Observation of the Topological {{Anderson}} Insulator in Disordered Atomic Wires},
  author = {Meier, Eric J. and An, Fangzhao Alex and Dauphin, Alexandre and Maffei, Maria and Massignan, Pietro and Hughes, Taylor L. and Gadway, Bryce},
  year = {2018},
  month = nov,
  journal = {Science},
  volume = {362},
  number = {6417},
  pages = {929--933},
}

@article{mondragon-shemTopologicalCriticalityChiralSymmetric2014,
  title = {Topological {{Criticality}} in the {{Chiral-Symmetric AIII Class}} at {{Strong Disorder}}},
  author = {{Mondragon-Shem}, Ian and Hughes, Taylor L. and Song, Juntao and Prodan, Emil},
  year = {2014},
  month = jul,
  journal = {Phys. Rev. Lett.},
  volume = {113},
  number = {4},
  pages = {046802},
  publisher = {American Physical Society},
}

@article{nashTopologicalMechanicsGyroscopic2015,
  title = {Topological Mechanics of Gyroscopic Metamaterials},
  author = {Nash, Lisa M. and Kleckner, Dustin and Read, Alismari and Vitelli, Vincenzo and Turner, Ari M. and Irvine, William T. M.},
  year = {2015},
  month = nov,
  journal = {Proceedings of the National Academy of Sciences},
  volume = {112},
  number = {47},
  pages = {14495--14500},
}

@article{niObservationHigherorderTopological2019,
  title = {Observation of Higher-Order Topological Acoustic States Protected by Generalized Chiral Symmetry},
  author = {Ni, Xiang and Weiner, Matthew and Al{\`u}, Andrea and Khanikaev, Alexander B.},
  year = {2019},
  month = feb,
  journal = {Nature Mater},
  volume = {18},
  number = {2},
  pages = {113--120},
}

@article{oudichEngineeredMoirePhotonic2024,
  title = {Engineered Moir{\'e} Photonic and Phononic Superlattices},
  author = {Oudich, Mourad and Kong, Xianghong and Zhang, Tan and Qiu, Chengwei and Jing, Yun},
  year = {2024},
  month = sep,
  journal = {Nat. Mater.},
  volume = {23},
  number = {9},
  pages = {1169--1178},
}

@article{pauloseTopologicalModesBound2015,
  title = {Topological Modes Bound to Dislocations in Mechanical Metamaterials},
  author = {Paulose, Jayson and Chen, Bryan Gin-ge and Vitelli, Vincenzo},
  year = {2015},
  month = feb,
  journal = {Nature Phys},
  volume = {11},
  number = {2},
  pages = {153--156},
}

@article{quTopologicalPhotonicAlloy2024,
  title = {Topological {{Photonic Alloy}}},
  author = {Qu, Tiantao and Wang, Mudi and Cheng, Xiaoyu and Cui, Xiaohan and Zhang, Ruo-Yang and Zhang, Zhao-Qing and Zhang, Lei and Chen, Jun and Chan, C. T.},
  year = {2024},
  month = may,
  journal = {Phys. Rev. Lett.},
  volume = {132},
  number = {22},
  pages = {223802},
}

@article{rechtsmanPhotonicFloquetTopological2013,
  title = {Photonic {{Floquet}} Topological Insulators},
  author = {Rechtsman, Mikael C. and Zeuner, Julia M. and Plotnik, Yonatan and Lumer, Yaakov and Podolsky, Daniel and Dreisow, Felix and Nolte, Stefan and Segev, Mordechai and Szameit, Alexander},
  year = {2013},
  month = apr,
  journal = {Nature},
  volume = {496},
  number = {7444},
  pages = {196--200},
}

@article{restaInsulatingStateMatter2011,
  title = {The Insulating State of Matter: A Geometrical Theory},
  shorttitle = {The Insulating State of Matter},
  author = {Resta, R.},
  year = {2011},
  month = jan,
  journal = {Eur. Phys. J. B},
  volume = {79},
  number = {2},
  pages = {121--137},
}

@article{rosendolopezFlatBandsMagicAngle2020,
  title = {Flat {{Bands}} in {{Magic-Angle Vibrating Plates}}},
  author = {Rosendo L{\'o}pez, Mar{\'i}a and Pe{\~n}aranda, Fernando and Christensen, Johan and {San-Jose}, Pablo},
  year = {2020},
  month = nov,
  journal = {Phys. Rev. Lett.},
  volume = {125},
  number = {21},
  pages = {214301},
}

@article{schaibleyValleytronics2DMaterials2016,
  title = {Valleytronics in {{2D}} Materials},
  author = {Schaibley, John R. and Yu, Hongyi and Clark, Genevieve and Rivera, Pasqual and Ross, Jason S. and Seyler, Kyle L. and Yao, Wang and Xu, Xiaodong},
  year = {2016},
  month = aug,
  journal = {Nat Rev Mater},
  volume = {1},
  number = {11},
  pages = {1--15},
}

@article{scheibnerOddElasticity2020a,
  title = {Odd Elasticity},
  author = {Scheibner, Colin and Souslov, Anton and Banerjee, Debarghya and Sur{\'o}wka, Piotr and Irvine, William T. M. and Vitelli, Vincenzo},
  year = {2020},
  month = apr,
  journal = {Nat. Phys.},
  volume = {16},
  number = {4},
  pages = {475--480},
}

@article{shiDisorderinducedTopologicalPhase2021,
  title = {Disorder-Induced Topological Phase Transition in a One-Dimensional Mechanical System},
  author = {Shi, Xiaotian and Kiorpelidis, Ioannis and Chaunsali, Rajesh and Achilleos, Vassos and Theocharis, Georgios and Yang, Jinkyu},
  year = {2021},
  month = jul,
  journal = {Phys. Rev. Res.},
  volume = {3},
  number = {3},
  pages = {033012},
}

@misc{shiTopologicalPhaseTransition2022,
  title = {Topological Phase Transition in Disordered Elastic Quantum Spin {{Hall}} System},
  author = {Shi, Xiaotian and Chaunsali, Rajesh and Theocharis, Georgios and Huang, Huaqing and Zhu, Rui and Yang, Jinkyu},
  year = {2022},
  month = dec,
  number = {arXiv:2212.09435},
  eprint = {2212.09435},
  primaryclass = {cond-mat, physics:physics},
}

@article{soneNonlinearityinducedTopologicalPhase2024,
  title = {Nonlinearity-Induced Topological Phase Transition Characterized by the Nonlinear {{Chern}} Number},
  author = {Sone, Kazuki and Ezawa, Motohiko and Ashida, Yuto and Yoshioka, Nobuyuki and Sagawa, Takahiro},
  year = {2024},
  month = apr,
  journal = {Nat. Phys.},
  pages = {1--7},
}

@article{songNonHermitianTopologicalInvariants2019,
  title = {Non-{{Hermitian Topological Invariants}} in {{Real Space}}},
  author = {Song, Fei and Yao, Shunyu and Wang, Zhong},
  year = {2019},
  month = dec,
  journal = {Phys. Rev. Lett.},
  volume = {123},
  number = {24},
  pages = {246801},
}

@article{sykesLocalTopologicalMarkers2021,
  title = {Local Topological Markers in Odd Dimensions},
  author = {Sykes, Joseph and Barnett, Ryan},
  year = {2021},
  month = apr,
  journal = {Phys. Rev. B},
  volume = {103},
  number = {15},
  pages = {155134},
}

@article{tuloupNonlinearityInducedTopological2020,
  title = {Nonlinearity Induced Topological Physics in Momentum Space and Real Space},
  author = {Tuloup, Thomas and Bomantara, Raditya Weda and Lee, Ching Hua and Gong, Jiangbin},
  year = {2020},
  month = sep,
  journal = {Phys. Rev. B},
  volume = {102},
  number = {11},
  pages = {115411},
}

@article{wangStructuralAmorphizationInducedTopological2022,
  title = {Structural {{Amorphization-Induced Topological Order}}},
  author = {Wang, Citian and Cheng, Ting and Liu, Zhirong and Liu, Feng and Huang, Huaqing},
  year = {2022},
  month = feb,
  journal = {Phys. Rev. Lett.},
  volume = {128},
  number = {5},
  pages = {056401},
}

@article{wuSchemeAchievingTopological2015,
  title = {Scheme for {{Achieving}} a {{Topological Photonic Crystal}} by {{Using Dielectric Material}}},
  author = {Wu, Long-Hua and Hu, Xiao},
  year = {2015},
  month = jun,
  journal = {Phys. Rev. Lett.},
  volume = {114},
  number = {22},
  pages = {223901},
}

@article{xiaoSyntheticGaugeFlux2015a,
  title = {Synthetic Gauge Flux and {{Weyl}} Points in Acoustic Systems},
  author = {Xiao, Meng and Chen, Wen-Jie and He, Wen-Yu and Chan, C. T.},
  year = {2015},
  month = nov,
  journal = {Nature Phys},
  volume = {11},
  number = {11},
  pages = {920--924},
}

@article{xueAcousticHigherorderTopological2019,
  title = {Acoustic Higher-Order Topological Insulator on a Kagome Lattice},
  author = {Xue, Haoran and Yang, Yahui and Gao, Fei and Chong, Yidong and Zhang, Baile},
  year = {2019},
  month = feb,
  journal = {Nature Mater},
  volume = {18},
  number = {2},
  pages = {108--112},
}

@article{xuHydrodynamicMoireSuperlattice2024,
  title = {Hydrodynamic Moir{\'e} Superlattice},
  author = {Xu, Guoqiang and Zhou, Xue and Chen, Weijin and Hu, Guangwei and Yan, Zhiyuan and Li, Zhipeng and Yang, Shuihua and Qiu, Cheng-Wei},
  year = {2024},
  month = dec,
  journal = {Science},
  volume = {386},
  number = {6728},
  pages = {1377--1383},
}

@article{xuObservationBulkQuadrupole2023,
  title = {Observation of Bulk Quadrupole in Topological Heat Transport},
  author = {Xu, Guoqiang and Zhou, Xue and Yang, Shuihua and Wu, Jing and Qiu, Cheng-Wei},
  year = {2023},
  month = jun,
  journal = {Nat Commun},
  volume = {14},
  number = {1},
  pages = {3252},
}

@article{yangAcousticTypeIIWeyl2016,
  title = {Acoustic {{Type-II Weyl Nodes}} from {{Stacking Dimerized Chains}}},
  author = {Yang, Zhaoju and Zhang, Baile},
  year = {2016},
  month = nov,
  journal = {Phys. Rev. Lett.},
  volume = {117},
  number = {22},
  pages = {224301},
}

@article{yangStrainInducedGaugeField2017,
  title = {Strain-{{Induced Gauge Field}} and {{Landau Levels}} in {{Acoustic Structures}}},
  author = {Yang, Zhaoju and Gao, Fei and Yang, Yahui and Zhang, Baile},
  year = {2017},
  month = may,
  journal = {Phys. Rev. Lett.},
  volume = {118},
  number = {19},
  pages = {194301},
}

@article{yangTopologicalAcoustics2015,
  title = {Topological {{Acoustics}}},
  author = {Yang, Zhaoju and Gao, Fei and Shi, Xihang and Lin, Xiao and Gao, Zhen and Chong, Yidong and Zhang, Baile},
  year = {2015},
  month = mar,
  journal = {Phys. Rev. Lett.},
  volume = {114},
  number = {11},
  pages = {114301},
}

@article{yanPseudomagneticFieldsEnabled2021,
  title = {Pseudomagnetic {{Fields Enabled Manipulation}} of {{On-Chip Elastic Waves}}},
  author = {Yan, Mou and Deng, Weiyin and Huang, Xueqin and Wu, Ying and Yang, Yating and Lu, Jiuyang and Li, Feng and Liu, Zhengyou},
  year = {2021},
  month = sep,
  journal = {Phys. Rev. Lett.},
  volume = {127},
  number = {13},
  pages = {136401},
}

@article{yanTopologicalMaterialsWeyl2017,
  title = {Topological {{Materials}}: {{Weyl Semimetals}}},
  shorttitle = {Topological {{Materials}}},
  author = {Yan, Binghai and Felser, Claudia},
  year = {2017},
  journal = {Annual Review of Condensed Matter Physics},
  volume = {8},
  number = {1},
  pages = {337--354},
}

@article{zhangProgrammableElasticValley2019,
  title = {Programmable Elastic Valley {{Hall}} Insulator with Tunable Interface Propagation Routes},
  author = {Zhang, Quan and Chen, Yi and Zhang, Kai and Hu, Gengkai},
  year = {2019},
  month = apr,
  journal = {Extreme Mechanics Letters},
  volume = {28},
  pages = {76--80},
  issn = {2352-4316},
}

@article{zhaoElasticValleySpin2022,
  title = {Elastic {{Valley Spin Controlled Chiral Coupling}} in {{Topological Valley Phononic Crystals}}},
  author = {Zhao, Jinfeng and Yang, Chenwen and Yuan, Weitao and Zhang, Danmei and Long, Yang and Pan, Yongdong and Chen, Hong and Zhong, Zheng and Ren, Jie},
  year = {2022},
  month = dec,
  journal = {Phys. Rev. Lett.},
  volume = {129},
  number = {27},
  pages = {275501},
}

@article{zhaoNonHermitianTopologicalLight2019,
  title = {Non-{{Hermitian}} Topological Light Steering},
  author = {Zhao, Han and Qiao, Xingdu and Wu, Tianwei and Midya, Bikashkali and Longhi, Stefano and Feng, Liang},
  year = {2019},
  month = sep,
  journal = {Science},
  volume = {365},
  number = {6458},
  pages = {1163--1166},
  issn = {0036-8075, 1095-9203},
}

@article{zhaoRealizationHaldaneChern2024,
  title = {Realization of the {{Haldane Chern}} Insulator in a Moir{\'e} Lattice},
  author = {Zhao, Wenjin and Kang, Kaifei and Zhang, Yichi and Kn{\"u}ppel, Patrick and Tao, Zui and Li, Lizhong and Tschirhart, Charles L. and Redekop, Evgeny and Watanabe, Kenji and Taniguchi, Takashi and Young, Andrea F. and Shan, Jie and Mak, Kin Fai},
  year = {2024},
  month = jan,
  journal = {Nat. Phys.},
  pages = {1--6},
}

@article{zhouTopologicalEdgeFloppy2018,
  title = {Topological {{Edge Floppy Modes}} in {{Disordered Fiber Networks}}},
  author = {Zhou, Di and Zhang, Leyou and Mao, Xiaoming},
  year = {2018},
  month = feb,
  journal = {Phys. Rev. Lett.},
  volume = {120},
  number = {6},
  pages = {068003},
}

@article{zhouTopologicalInvariantAnomalous2022a,
  title = {Topological Invariant and Anomalous Edge Modes of Strongly Nonlinear Systems},
  author = {Zhou, Di and Rocklin, D. Zeb and Leamy, Michael and Yao, Yugui},
  year = {2022},
  month = jun,
  journal = {Nat Commun},
  volume = {13},
  number = {1},
  pages = {3379},
}

@article{zhao_elastic_2022,
  title={Elastic Valley Spin Controlled Chiral Coupling in Topological Valley Phononic Crystals},
  author={Zhao, Jinfeng and Yang, Chenwen and Yuan, Weitao and Zhang, Danmei and Long, Yang and Pan, Yongdong and Chen, Hong and Zhong, Zheng and Ren, Jie},
  journal={Physical Review Letters},
  volume={129},
  number={27},
  pages={275501},
  year={2022},
  publisher={APS}
}

@article{liu_tunable_2018,
  title={Tunable acoustic valley--hall edge states in reconfigurable phononic elastic waveguides},
  author={Liu, Ting-Wei and Semperlotti, Fabio},
  journal={Physical Review Applied},
  volume={9},
  number={1},
  pages={014001},
  year={2018},
  publisher={APS}
}

@article{wang2024locally,
  title={Locally strained 2D materials: preparation, properties, and applications},
  author={Wang, Jingwei and He, Liqiong and Zhang, Yunhao and Nong, Huiyu and Li, Shengnan and Wu, Qinke and Tan, Junyang and Liu, Bilu},
  journal={Advanced Materials},
  volume={36},
  number={23},
  pages={2314145},
  year={2024},
  publisher={Wiley Online Library}
}

@article{forsythe2018band,
  title={Band structure engineering of 2D materials using patterned dielectric superlattices},
  author={Forsythe, Carlos and Zhou, Xiaodong and Watanabe, Kenji and Taniguchi, Takashi and Pasupathy, Abhay and Moon, Pilkyung and Koshino, Mikito and Kim, Philip and Dean, Cory R},
  journal={Nature nanotechnology},
  volume={13},
  number={7},
  pages={566--571},
  year={2018},
  publisher={Nature Publishing Group UK London}
}

@article{nie2025designing,
  title={Designing unique mechanical modes through an extension of the quantum hopping method},
  author={Nie, Haoran and Shen, Xiangying and Xu, Lei},
  journal={Proceedings of the National Academy of Sciences},
  volume={122},
  number={46},
  pages={e2423603122},
  year={2025},
  publisher={National Academy of Sciences}
}

\end{document}


\title{\fontsize{19}{46}\selectfont Supplemental Materials}
\author{Haoran Nie}
\author{Chaoran Jiang}
\author{Xiangying Shen}
\author{Lei Xu}
\maketitle
\subsection{\label{sec:level1}I. Construction of mechanical network}

Mechanical system stands different with quantum system for the difference between Newton equation and Schrodinger equations.However, they share some similarities when studying on mechanical vibrational systems, where the eigen equation reads: $Du=-\omega^2u$. The quantum Hamiltonian, based on tight-binding model, can be quickly written and analyzed with hopping method, where the transition probability between sites can be treated as hopping factors. This, however, becomes different when writing mechanical dynamic matrix. Since forces in classical mechanics cannot exist independently, any mechanical hopping action always corresponds to the coordinates of two nodes, and the corresponding term is: $u_i^\dagger t_{ij}(u_j-u_i)$ for node $i$ and $u_j^\dagger t_{ij}(u_i-u_j)$ for node $j$, where $t_{ij}$ is equivalent spring stiffness between i and j. That leads to additional onsite diagonal terms $-t_{ij}$ on the diagonal terms $D_{ii},D_{jj}$. 

\begin{figure}[htbp]
\begin{center}
\centerline{\includegraphics[width=0.5\linewidth]{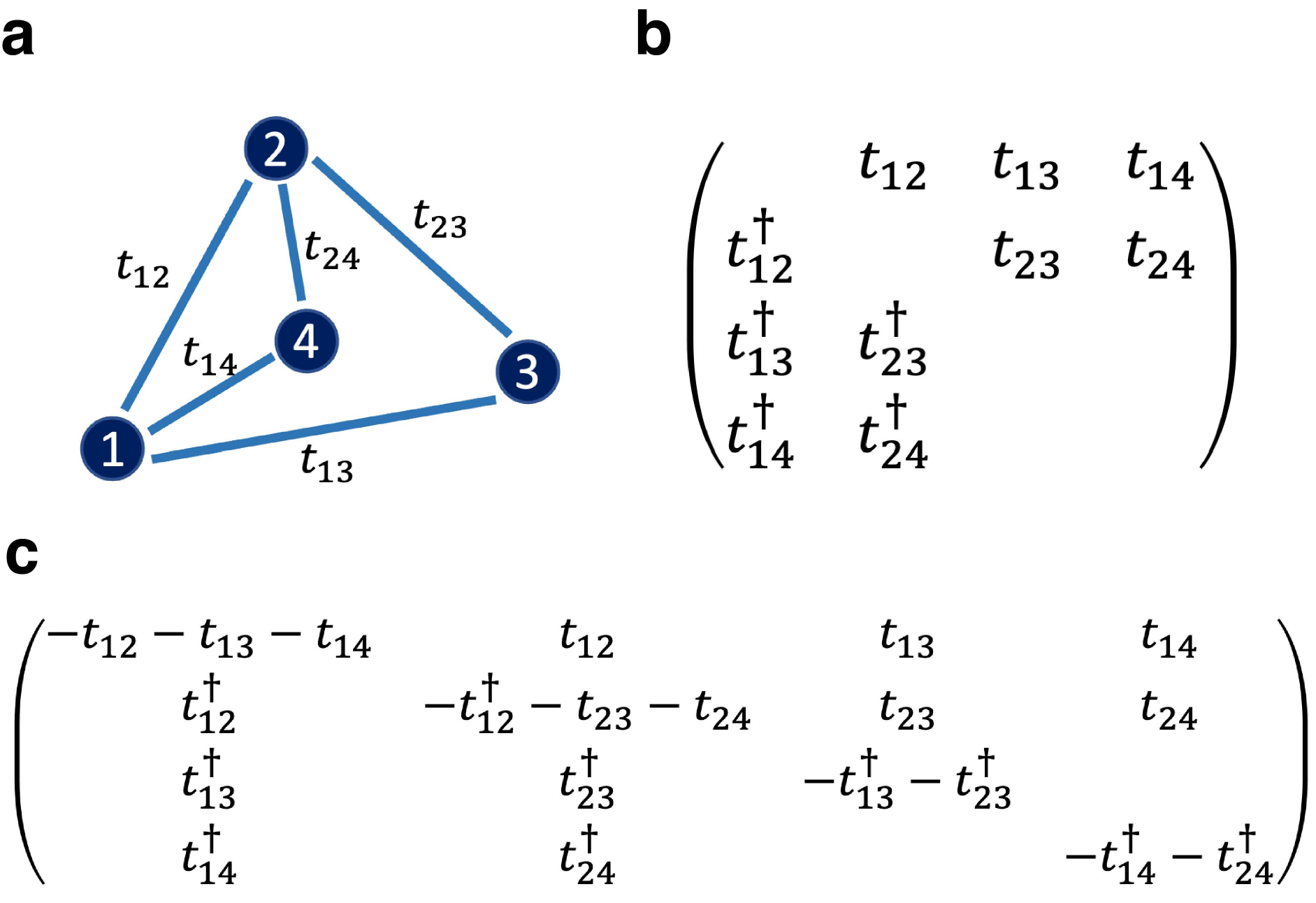}}
\begin{flushleft}\label{s1} FIG. S1: (a) A schematic of network structure with corresponding hopping factors. (b) A simple example of tight-binding Hamiltonian. (c) An example of mechanical dynamic matrix.\end{flushleft}
\vspace{-0cm}
\end{center}
\end{figure}

We take a very simple example shown in FIG. S1, where we choose a 4 sites network as exhibited in (a) and the coupling relation between $i,j$ is marked by $t_{ij}$. We don't consider other complicated situations such as spins. The quantum Hamiltonian can be written in (b) where the dynamic matrix should be constructed in (c). The hopping factors can be change into sub-matrix if the system process complicate interactions (such as spin). The biggest difference is that the diagonal terms are linked with off-diagonal terms in mechanical networks. This in fact reduces degree of freedom in the system. Nevertheless, the diagonal terms in eigen-equations, which are supposed to represent the energy levels on each sites in quantum Hamiltonian, can be re-constructed in mechanical dynamic matrix by introducing separate springs connected with fixed anchor points. This kinds of connections only contribute to diagonal terms in the dynamic matrix, decouple with the off-diagonal terms and introduce equivalent mechanical onsite energy levels/potentials. By manipulating each of the onsite mechanical onsite potential as a function of real space index (i,j), we can introduce an equivalent regulating field similar as electric field on graphene. It is worth mentioning that the mechanical hopping factor $t_{ij}$ here does not necessarily represent the stiffness of the springs prior to the lattice but rather indicates the strength of the connections between the lattice sites for the propagation of mechanical modes. For more details, please refer to SI Section V.

\subsection{\label{sec:level2}II. Characterization of topological changes in valley}

As illustrated in the main text, gradually increasing the regulating field may twist the edge band to cross the band gap. And the phase angles' distribution accumulates at the valley area. 
\begin{equation*}
    \delta \theta(n)=Arg(\Pi_{p}\langle u(k_{n,p})|u(k_{n,p+1})\rangle)
\end{equation*}
However, this is a necessary but not sufficient condition for determining system's topology. Because the overall topological phase angle of the band in topological materials utilizing the local valley topology in k-space is often 0, and there is not a strict method for calculating the phase within the valley region, the results can be significantly influenced by the choice of the region. Therefore, we introduced the formulas in the main text to explore the topological properties of the system from a new perspective:
\begin{equation}\label{eq2}
    \mathcal{C}(x)=\mathop{\rm{average}}\limits_{\alpha,\beta}(Q_{\alpha\beta}[Q_{\beta\alpha},X]_x+Q_{\beta\alpha}[X,Q_{\alpha\beta}]_x)
\end{equation}
where $\alpha,\beta$ are sub-lattice notations; $X$ is position operator; $Q_{\alpha\beta}=\Gamma_\alpha(P_+-P_-)\Gamma_\beta$ and the same for $Q_{\beta\alpha}$; $\Gamma$ is sub-lattice projectors; $P_\pm$ is band projectors and $\pm$ means above or below the gap. The sub-lattice operator $\Gamma_\alpha,\Gamma_\beta$ are diagonal matrix with corresponding sub-lattice positions equal 1 and others are 0. The position operator $X$ is also a diagonal matrix with diagonal terms equal the index of unite cells. $P_+,P_-$ are the matrices generated by exterior product of the eigenvectors corresponding to the eigenvalues near the spectrum gap (the vibrational excitation does not fill the energy band as an electronic system does, only the nearby energy band structure matters), representing the upper and lower states, respectively. Because mechanical vibrations do not possess the complete filling and vacancy bands characteristic of electronic fermionic systems, the results of the Lattice Topological Model (LTM) can take continuous values.

\subsection{\label{sec:level3}III. Effective mass $m_{\mathrm{eff}}$ influenced by ESP}

Building on our effective mass model for ESP-regulated valley topological states in the main text, we show that the ESP alters the spatial distribution of $m_\text{eff}$ along the x-axis via modification of the Hamiltonian. Upon the ESP-induced transition from an insulator to a topological conductor, the position of the in-gap conducting state aligns with the sign reversal of $m_\text{eff}$. We discuss this behavior herein, and use a lattice polarizability model to visualize the spatial variation of $m_\text{eff}$.

\noindent\textbf{1. Near gap two-band description.}

Near the valley momenta $k_y \simeq K_y$ (where the bulk gap is minimal), the two bands closest to the gap are well captured by a two-component pseudospin basis associated with the two sublattices $(A,B)$. In this subspace, the dynamics can be approximated by an effective Dirac-type model,

\begin{equation*}\label{eq3}
    H(\bm{r},\bm{k})=v(\tau k_x \sigma_x + k_y \sigma_y)+m_{\text{eff}}(\bm{r},k_y)\sigma_z + \bm{V(\bm{r})}
    = v\Big(\tau(-i\partial_x)\sigma_x + (k_y-K_y)\sigma_y\Big)
    + m_{\mathrm{eff}}(x,k_y)\,\sigma_z
    +\bm{V(x)}
\end{equation*}
where $\tau=\pm 1$ labels the two valleys, $v$ is the effective velocity, $\sigma_{x,y,z}$ are Pauli matrices acting on the $(A,B)$ sublattice, and $V_{\mathrm{eff}}$ shifts the local spectral center. The quantity
$m_{\mathrm{eff}}$ plays the role of effective mass that controls the local gap and its topological character. Microscopically, the external elastic potential (ESP) enters the stripe matrix as a sublattice-dependent onsite term,
\begin{equation*}
D(k_y)=D_0(k_y)+ p(x)W
\end{equation*}
where $p(x)$ is a slowly varying ESP profile along $x$, and $W$ encodes the (generally different) coupling of the ESP to the two sublattices (e.g., with strengths $p_1$ and $p_2$ in our implementation).
In the full stripe spectrum, the ESP also induces band mixing between the two low-energy modes and remote bands (and, in finite geometries, boundary-quantized modes). Integrating out these higher-energy degrees of freedom yields a second-order self-energy correction in the low-energy subspace. Formally, using a Schrieffer--Wolff (SW) projection with $P$ the low-energy projector and $Q=1-P$,
\begin{equation*}
\delta H^{(2)}(x)\approx
-\,P\,H_p(x)\,Q\;\frac{1}{H_Q-E}\;Q\,H_p(x)\,P
\qquad
H_p(x)=p(x)\,W,
\end{equation*}
which generically produces a renormalization of the effective mass term in the Dirac expansion. Consequently, the renormalized mass can be parameterized as
\begin{equation*}
m_{\mathrm{eff}}(x,k_y)\approx
m_0+\eta\,p(x)-\lambda\,p(x)^2+\beta\,(k_y-K_y)^2
\end{equation*}
where $m_0$ is the baseline mass (half the valley gap in the absence of ESP), $\eta$ is the linear response to ESP, and the quadratic term $-\lambda p^2$ encodes the SW/self-energy renormalization. The coefficient $\beta$ accounts for the leading momentum dependence near the valley. In addition, the finite boundary conditions reshapes the eigenvector basis of the system, which leads to a coupling between Pauli matrix components and in turn modulates the effective mass $m_\text{eff}$. It explains why an ESP can drive the system through a local inversion of $m_\text{eff}$.

\begin{figure}[htbp]
\begin{center}
\centerline{\includegraphics[width=0.7\linewidth]{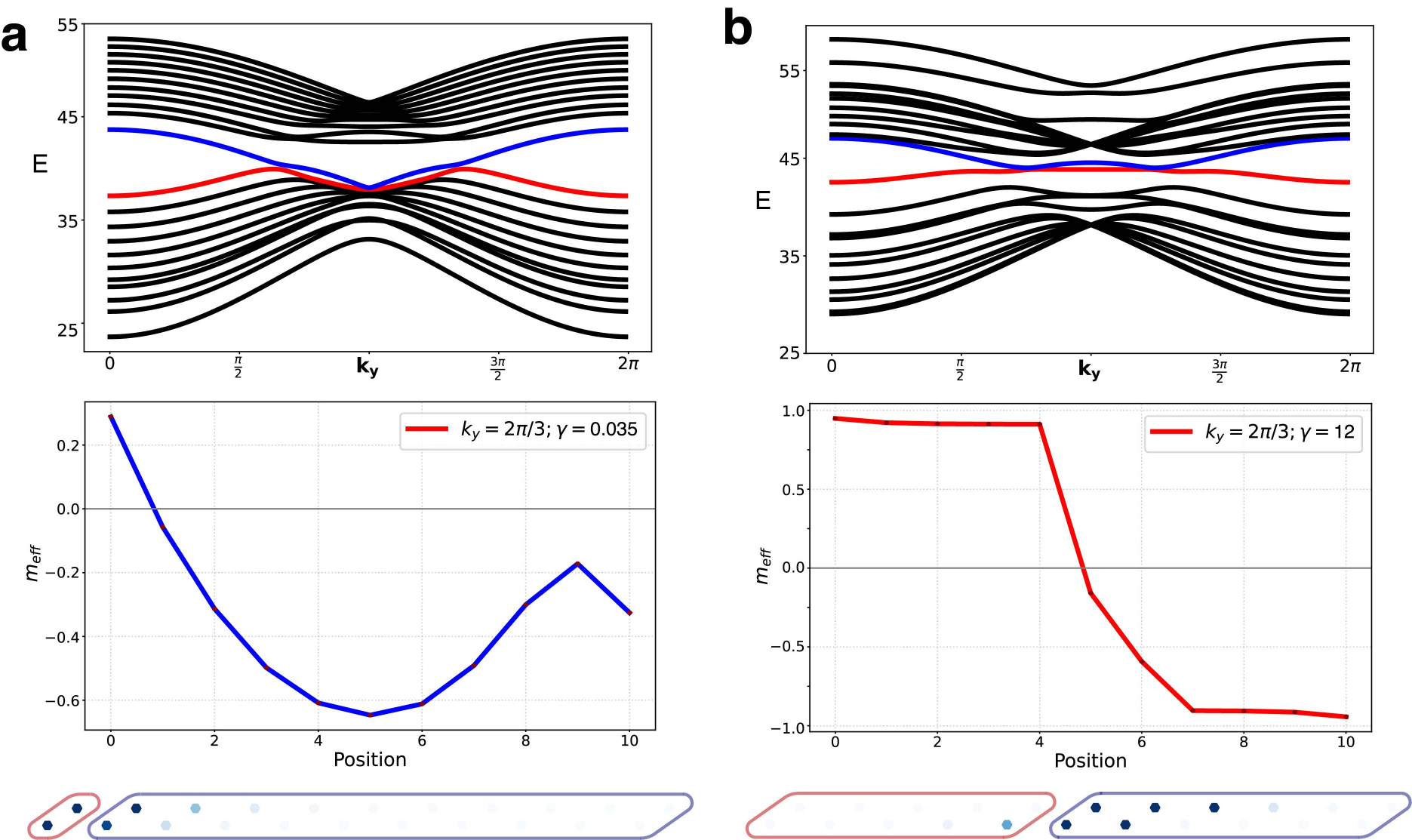}}
\begin{flushleft}\label{s2} FIG. S2: Comparison of the band maps (top panels), effective valley masses $m_\text{eff}$ (middle panels), and corresponding modes(bottom panels). (a) is quadratic-shape potential with $\gamma=0.035$. (b) is Gaussian-shape potential with $\gamma=12$. (a) middle and bottom panels plot the $m_\text{eff}$ and mode of the blue band in the up panel, while in (b) they correspond to the red band. The circled colors of the bottom modes corresponds to different $m_\text{eff}$ signs.
 \end{flushleft}
\vspace{-0cm}
\end{center}
\end{figure}

\noindent\textbf{2. Role of $m_{\text{eff}}$: domain walls and in-gap modes}

A minimal description of a domain-wall mode is given by the 1D Dirac Hamiltonian:
\begin{equation*}
H_{\mathrm{JR}}=
- i v\,\sigma_x \partial_x + m(x)\,\sigma_z
\end{equation*}
where $v$ is the Dirac velocity, $\sigma_{x,z}$ are Pauli matrices acting on an internal two-component basis, and $m(x)$ is a spatially varying Dirac mass. The bulk spectrum for constant $m$ is gapped:
\begin{equation*}
E(k)=\pm \sqrt{(v k)^2+m^2}
\end{equation*}
with a gap size $2|m|$.

A mass inversion (domain wall) occurs when $m(x)$ changes sign at some position $x_0$, i.e.,
\begin{equation*}
\mathrm{sgn}\!\big(m(x<x_0)\big)\neq \mathrm{sgn}\!\big(m(x>x_0)\big)
\end{equation*}

In this case, the Dirac equation $H\psi=E\psi$ admits a localized in-gap solution. In particular, at $E=0$ (or other reference energy) one obtains a normalizable bound state of the form:

\begin{equation*}
\psi_0(x)\propto
\exp\!\Bigg[-\frac{1}{v}\int_{x_0}^{x} m(x')\,dx'\Bigg]\,
\chi
\end{equation*}

where $\chi$ is a constant spinor satisfying a fixed pseudospin polarization condition (equivalently, $\sigma_y \chi = \pm \chi$ depending on conventions). The exponential factor shows that the wavefunction is confined near the domain wall, with a characteristic localization length set by the mass gradient; for a sharp kink $m(x)\approx m_0\,\mathrm{sgn}(x-x_0)$ one finds
\begin{equation*}
\xi \sim \frac{v}{|m_0|}
\end{equation*}

Physically, the mass sign change corresponds to a band inversion between two gapped phases. The bound state is protected by the change of the mass sign (and the associated change of the 1D topological index in the valley Hall materials as depicted in the main text), and it disappears when the sign inversion condition of mass inversion is not satisfied.

\noindent\textbf{3. Practical numerical presentation of effective mass}

Precise analytical derivation of the effective mass in realistic systems presents substantial difficulties. To address this, we introduce a straightforward approach to visualize the effective mass $m_\text{eff}$ reversal by computing the sublattice polarization. This quantity serves as a proxy for the $\sigma_z$ coefficient, thereby reflecting the effective mass. The sublattice (pseudospin) polarization provides a convenient indicator of the local mass sign. For an eigenmode $\mathbf{u}$, we define the cell-resolved sublattice polarization:
\begin{equation}\label{eq2}
P(x)\equiv \frac{|u_A(x)|^2-|u_B(x)|^2}{|u_A(x)|^2+|u_B(x)|^2}
\end{equation}
which estimates $\langle \sigma_z\rangle$ in the effective Dirac basis. While $P(x)$ is not identical to $m_{\text{eff}}(x)$ in general (it is normalized by the total pseudospin length and is affected by kinetic/momentum terms), the zero-crossing and spatial structure of $P(x)$ closely track the formation of mass domain walls in the parameter regime where the two-band Dirac description applies. 

To validate the method described in Eq. 2, which uses sublattice polarization to reflect the effective mass $m_\text{eff}$, we performed calculations for the spatially quadratic and Gaussian variation cases discussed in the main text; the results are presented in Fig. S2. When the ESP is applied to the hexagonal lattice, inducing topological conducting states (as shown in the upper panels of Fig. S2), we calculated the sublattice polarization ($\sim m_\text{eff}$) for these specific states --the blue band S2(a) and the red band in panel S2(b)-- within the valley region of the Brillouin zone. The resulting spatial distributions are displayed in the middle panels. Notably, zero-crossings of the Jackiw-Rebbi effective mass term are observed at the locations corresponding to the ESP-induced topological states (on the left side for the quadratic ESP and in the center for the Gaussian ESP). This indicates an inversion of the valley topological index across the zero-crossing point, which consequently gives rise to valley-protected topological states, consistent with our analytical predictions. In the bottom panels of Fig. S2, we plot the mode distributions of the in-gap topological conducting states, using differently colored frames to distinguish regions based on the sign of the sublattice polarization. This allows for a direct visual comparison between the mode distribution and the zero-crossing of the effective mass. 

In summary, the sign of the sublattice polarization provides a convenient approximation for describing $m_\text{eff}$, yielding results that correspond well with the two potential profiles in the main text and align with theoretical expectations.

\subsection{\label{sec:level4}IV. Examples of other regulating potential field's effects}

As demonstrated in the main text, the effects of the regulating potential on the quasi-one-dimensional chain system are highly diverse. Various types of regulating potentials can influence the band structure and topological properties of the system. This section provides additional examples of different potential types, showcasing the topological effects caused by other potential functions as a supplement to the content in the main text.

\begin{figure}[htbp]
\begin{center}
\centerline{\includegraphics[width=1.05\linewidth]{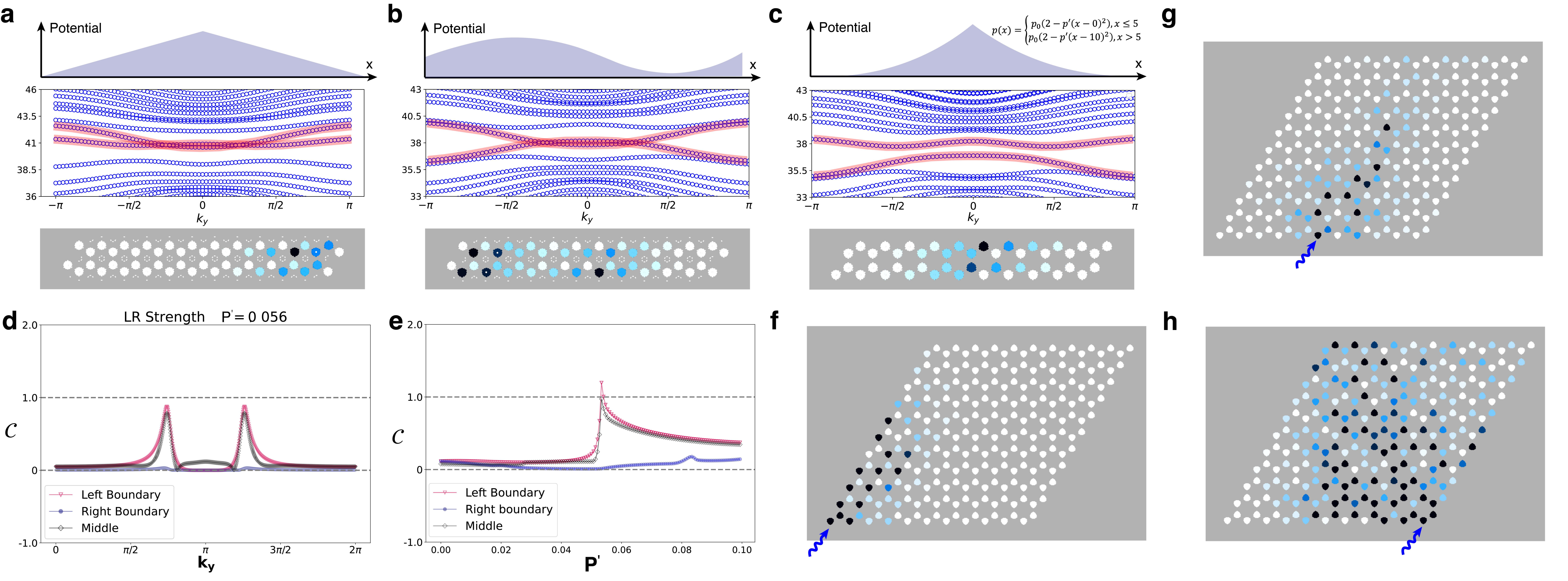}}
\begin{flushleft}\label{s3} FIG. S3: Some examples of other types of regulating potentials' effects. (a-c) are results of potential schematics (up panels), eigen-frequency spectra (middle panels) and in-gap mode of a quasi-1D chain (bottom panels) of reversed 'V'-shape (a), 'sinusoidal'-shape (b) and two-concatenated quadratic shape (c). (d.e) exhibits the LTM of two-concatenated quadratic shape potential. (d) shows the LTM in the 1st Brillouin zone at critical regulating potential strength while (e) exhibits the valley LTM varies with the change of potential strength. (f-h) are time zone excitation of the concatenated quadratic regulating potential of left geometric boundary (f), middle boundary (g) and right geometric boundary (h); the modes are recorded when t=3s. \end{flushleft}
\vspace{-0cm}
\end{center}
\end{figure}

After that, we apply V-shape, sinusoidal-shape, and two-concatenated quadratic shape regulating potentials to the quasi-one-dimensional chain system. The results are shown in Fig. S3 (a, b, c). The up panels are illustrations of the shapes of the applied regulating potentials. The middle panels are dispersion relations near the band gap in k-space after applying the regulating potentials, with boundary states highlighted in light red. The bottom panels exhibit the mode corresponding to one of the boundary states. It is evident that all three types of potentials influence the structure near the energy gap. In Fig. S3(a), the dispersion diagram shows two boundary states, each corresponding to the left and right geometric boundaries of the system. In Fig. S3(b), the boundary state in the dispersion diagram is trivial. In Fig. S3(c), the two boundary states correspond to the left geometric boundary and the center of the bulk, respectively (similar to the Gaussian potential discussed in the main text).

Furthermore, we calculated the Lattice Topological Model (LTM) distribution within the first Brillouin zone under the critical potential strength corresponding to the two-concatenated quadratic potential (Fig. S3 (d)) and the curve showing the average LTM in the valley region with the increase of potential strength (Fig. S3 (e)). It can be observed that under the two-concatenated quadratic potential field, the middle boundary (at the junction of the two quadratic potentials) and the left boundary exhibit nontrivial LTM, indicating that the propagating states within the band gap at these two boundaries possess a certain degree of topological protection. In contrast, the right boundary does not exhibit such protection. To verify this phenomenon, we performed simulations by exciting the left, middle, and right boundaries at corresponding frequencies around the gap and analyzed the mode distributions at t=3s. The results are shown in Fig. S3 (f, g, h), respectively. For the left boundary and the middle boundary, the vibrational modes propagate forward without diffusing into the bulk, demonstrating the topological protection of the states. For the right boundary, the excitation is quickly scattered, indicating the absence of topological protection. These observations are consistent with the theoretical calculations, further confirming the conclusions.

\begin{figure}[htbp]
\begin{center}
\centerline{\includegraphics[width=0.7\linewidth]{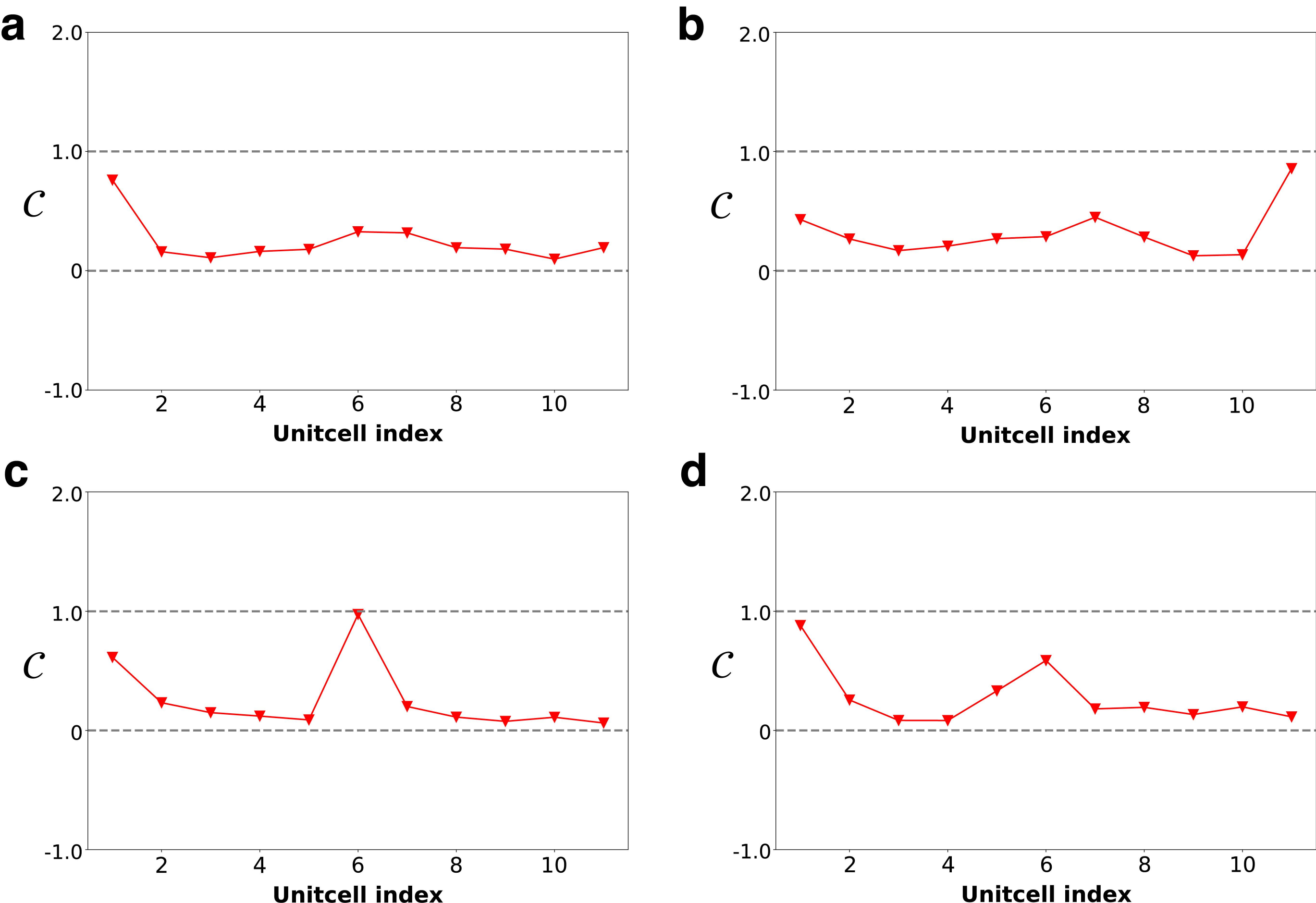}}
\begin{flushleft}
\label{s4}FIG. S4: The real-space LTM distribution of quadratic potential in the main text (a), reversed quadratic potential (b), Gauss potential in the main text (c) and the two-concatenated quadratic potential (d); the figures are taken under critical regulating potentials; the x-axis indicates unit-cell position. 
\end{flushleft}
\vspace{-0cm}
\end{center}
\end{figure}

As a supplement, we also calculated the real-space LTM distributions for several representative cases when the regulating potential field strength reaches its critical value. The results are shown in Fig. S4: (a) shows the real-space LTM distribution under the quadratic potential, as discussed in the main text. (b) exhibits the LTM distribution under the same potential field as in (a), but with the sub-lattice type reversed. (c) corresponds to the real-space LTM distribution under the Gaussian potential discussed in the main text.
(d) corresponds to the LTM distribution under the two-concatenated quadratic potential field shown in Fig. S3. It can be observed that the real-space LTM values are significantly larger at the nontrivial boundaries we mentioned, while other regions only exhibit relatively small values. As a supplement to the main text, these results align with our theoretical predictions and conclusions.

\subsection{\label{sec:level5}V. Topological effect on geometric boundary}

In the calculations and experiments on the Gaussian potential discussed in the main text, we observed that the Gaussian potential peak at the center of the bulk not only induces a nontrivial topological boundary state at the middle of the system but also influences the topological properties of the left boundary state. Interestingly, this influence exhibits a distant effect, as the Gaussian potential has nearly no direct impact on the left boundary due to its spatial distribution. To further explore this phenomenon, we calculated how far this distant effect can extend. It is reasonable to hypothesize that when the Gaussian regulating potential is fixed at the center of the quasi-1D chain, the remote effect will decay as the system size increases. In the limit of infinite system size, the lattice at the left boundary can be considered unaffected by the regulating potential, resembling a system without any potential field.

\begin{figure}[htbp]
\begin{center}
\centerline{\includegraphics[width=1.0\linewidth]{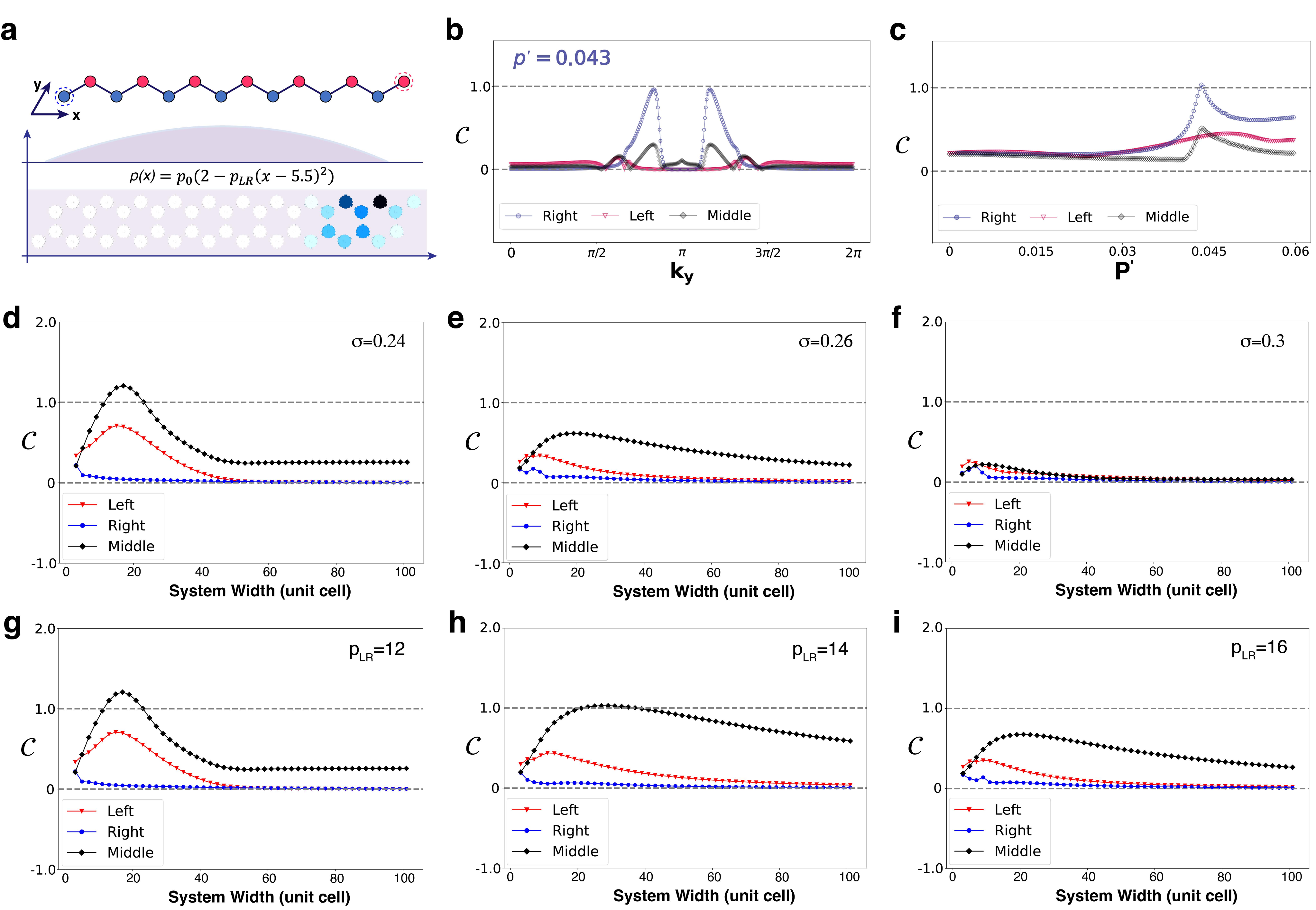}}
\begin{flushleft}\label{s5}FIG. S5: (a-c) are topological effect of different locations under quadratic potential, which is same as Fig. 3(b) in the main text but with sub-lattice reversed. (a) is the real-space structure, potential shape and simulation modes of the structure; (b) is the k-space LTM distribution at critical potential strength; (c) is the LTM of valley region with the increase of potential field. The other figures are the remote topological effect of Gauss regulating potential field with the change of lattice size. Where figures (d-f) are LTM variations under different Gaussian broadening $\sigma$ when $p'=12$; (g-i) are LTM variations under different potential amplitude $p'$ when $\sigma=0.24$. Colors red, blue, and black denote the left, right and middle boundaries, respectively. \end{flushleft}
\vspace{-0cm}
\end{center}
\end{figure}

In Fig. S5, we investigate the topological effect on geometric boundaries. Figs. (a-c) are results of quadratic potential field, which is same as Fig. 3(b) in the main text but with sub-lattice point reversed. Figs. (a–c) serve as a supplement to the main text. In the main text, under the same potential field, the nontrivial edge states are only located on the left boundary of the system. However, under the same potential field, by merely reversing the sub-lattice type, the nontrivial edge states shift to the right boundary of the system. These results, together with Fig. 4 and 5 in the main text, complement the discussion on the geometric boundary topological effects.

Furthermore, we investigate the dependence of the left boundary LTM on different Gaussian potential parameters. Fig. S5 (d-f) show the variation of the left boundary LTM with the change of lattice width under a Gaussian potential with a fixed peak strength of $p' = 12$ and different Gaussian widths.

It can be observed that as the Gaussian peak's width increases, the LTM values at both the left boundary and the middle boundary decrease. This is because a broader Gaussian peak results in a smoother potential distribution across the lattice, which increasingly resembles a uniform potential shift. Consequently, its impact on the band topology becomes weaker. On the other hand, when the Gaussian width is too narrow, the regulating field barely affects the surrounding unit cells, leading to a trivial effect on the topology. 

Fig. S5 (g-i) illustrate the variation of the LTM at the three boundaries (left, middle, and right) with the change of lattice width under a fixed Gaussian width ($\sigma=0.24$) but with different potential peak strengths. At higher potential strengths, the LTM values decay more slowly with increasing lattice size, indicating that stronger potentials maintain their influence over longer distances. However, excessively increasing the regulating potential strength does not always yield positive effects. When the potential becomes too strong, its impact on the topology diminishes.
The reason for this phenomenon is that overly strong regulating potentials cause the onsite potential strength at each lattice point to dominate over the inter-site coupling. In this regime, the lattice interactions become negligible, and the system transitions to a localized state governed primarily by the onsite potential. As a result, the topological properties weaken, which is an expected outcome.

These findings further clarify the balance required in regulating potential parameters to achieve and maintain nontrivial topological effects.

\subsection{\label{sec:level6}VI. Construction of experimental setups}

According to the theory outlined in Section I, the degrees of freedom for the mechanical vibration lattice points are actually equal to the order of the hopping factor matrix elements in mechanical hopping. For example, in a 2D vibrating system, the hopping factor $t_{ij}$ is a 
2×2 matrix, corresponding to the x and y directions. Similarly, applying an onsite potential in this system also requires a 2×2 matrix to act on these two states. To simplify the system for greater clarity and ease of adjusting the onsite potential, we have creatively designed an out-of-plane vibrating system. In this system, we only observe the vibrational modes in the z direction. This design choice allows for a more straightforward analysis and manipulation of the onsite potential while maintaining a clear understanding of the system's dynamics.

To enhance the tunability of the coupling between lattice points, we designed a stretching system that exists under tension in the plane. In this system, all springs connecting the coupling lattice points are in a stretched state, with boundary lattice points connected to fixed points in the plane to balance all initial stresses. We assume that the initial distance between points $i,j$ is $l$, and the initial stress on the spring is $F_0$ Under small vibrations $u$ in the z direction, the coupling between points $i,j$ can be expressed as:
\begin{equation*}
    F_{ji}=F_0 \sin{\theta}\approx F_0\theta=F_0(u_j-u_i)/l=t_{ij}(u_j-u_i)
\end{equation*}
\begin{equation*}
        t_{ij}=F_0/l=\frac{k\Delta l}{l_0+\Delta l}\propto1/l
\end{equation*}
where $l_0$ is the effective length of springs and $k$ is the effective stiffness of springs, $\theta$ denotes the angle between $\vec{ij}$ and the horizontal plane. The above formula provides us with significant tunability. In addition to adjusting the stiffness of the springs, we can achieve different mechanical hopping factors by varying the initial stress and the lengths of the springs between lattice points. Even if the initial stresses used in the lattice need to be balanced, we can create an unbalanced hopping factor by changing the connection lengths. This flexibility allows for a more versatile manipulation of the system's mechanical properties, enabling a tailored approach to studying various dynamic behaviors. As for the onsite potential of each lattice point, we can directly achieve the necessary adjustments by using springs connected to fixed points along the z direction. In this case, the spring constant directly corresponds to the onsite potential. This relationship allows for straightforward control over the onsite potential, facilitating precise tuning of the system's behavior and properties. 

\begin{figure}[htbp]
\begin{center}
\centerline{\includegraphics[width=0.7\linewidth]{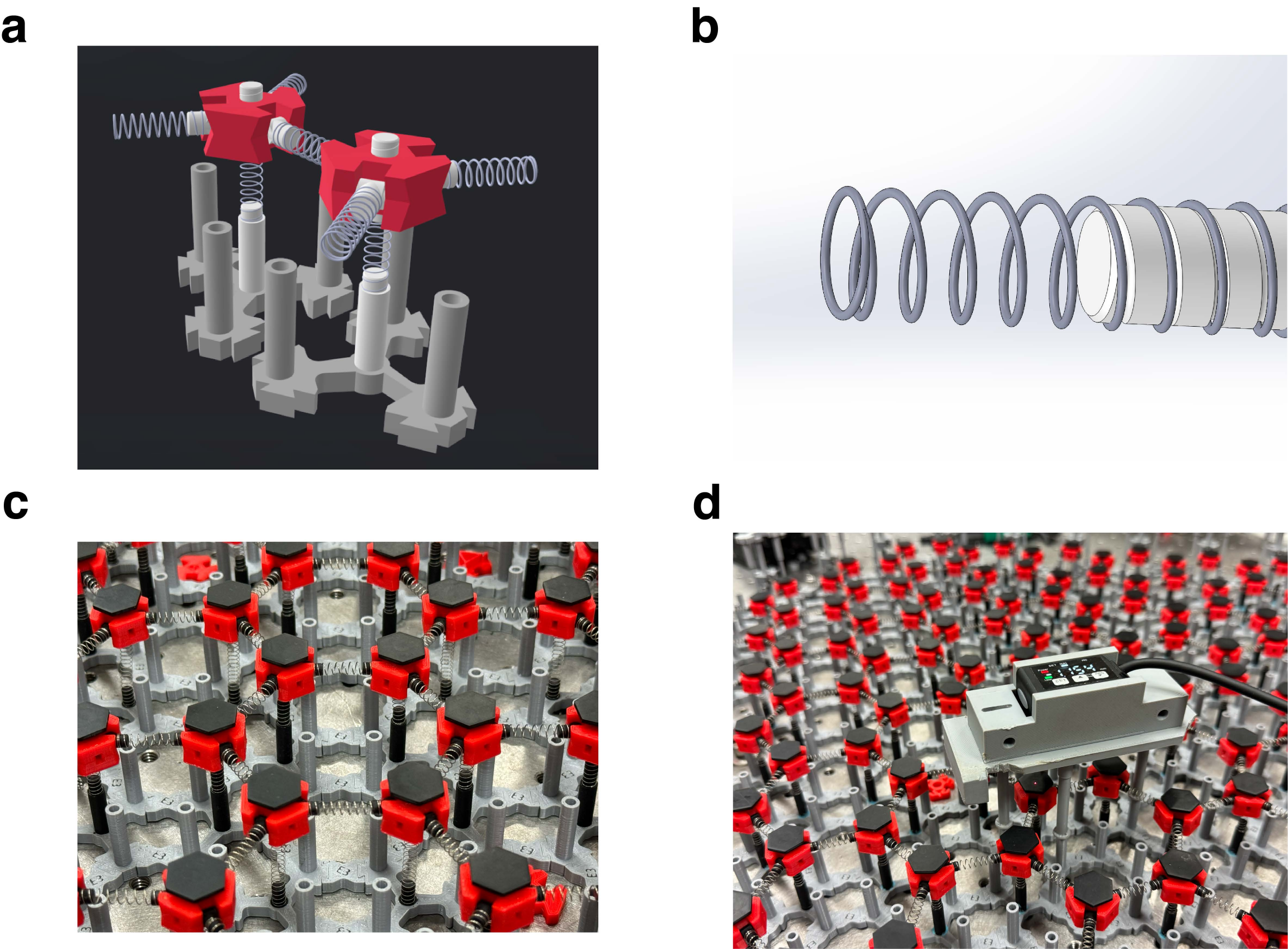}}
\begin{flushleft}\label{s6}FIG. S6: Schematics of experimental structures. (a) exhibits a rendered 3D structure of one unit cell in our system. (b) illustrates a schematic diagram showing how to adjust the effective length of the spring by using grooves that match the spring's pitch and diameter, thereby modifying the spring's stiffness. (c) presents a structural diagram of several unit cells in our experimental setup. We have additionally fixed a black resin cover plate on top of each lattice point to enhance the reflective properties, thereby improving the laser displacement measurement accuracy. (d) shows a schematic photo of the laser sensor measuring the displacement at a single lattice point. \end{flushleft}
\vspace{-0cm}
\end{center}
\end{figure}

Based on the aforementioned structural foundation, we designed a vibrating system consisting of 11 x 11 unit cells. The onsite regulating potential of the vibrating system can be expressed as: $f_{\alpha,\beta}(x)=p_{\alpha,\beta}p(x)$, where $\alpha,\beta$ denote the sub-lattice type and $p(x)$ is the real-space regulating function as shown in the main text. Originally, in order to open a gap of the eigenvalue spectrum, we take $p_\alpha=\rm{42.67N/m}$, $p_\beta=\rm{30.02N/m}$. The original in plane spring is $0.3\times 4\times 15$mm and the original stretch length is 3mm and the equivalent coupling between sites is $\rm{30N/m}$. We present our experimental and measurement structure in Fig. S6(a) The figure shows a rendered image of one unit cell. The initial stress in the experimental structure is controlled by adding a fixed bracket at the bottom to adjust the distance between the bases. (b) This illustration depicts how we use a light-curing printer to construct grooves that match the springs' pitch and wire diameter to adjust the effective stiffness. Our light-curing printer achieves micron-level precision, allowing for accurate matching of spring structures with a wire diameter of 0.3 mm. (c) This image is a close-up photograph of our actual experimental structure. In the actual measurements, we added a smooth black nylon cover plate over the PLA resin to better reflect the measurement light, enhancing the accuracy of the experiment. (d) This diagram illustrates the use of a laser displacement sensor to measure the vibrations of the lattice points. The sensor is fixed on a 3D-printed designed base to ensure stability and maintain the optimal measurement distance from the lattice points.

In the actual experimental measurements, we need to simultaneously measure multiple points to obtain the phase relationships between different lattice points (at least two points, one being a reference point). We present two measurement results in Fig. S7. (a) This figure shows the time-domain vibration curves of the measurement point and the reference point under a 39.9 Hz vibration excitation. (b) This plot displays the frequency distribution spectrum obtained from the Fourier transform of the two curves. It is evident that the vibration measurement is effective, and the spectrum peak after the Fourier transform aligns with the input vibration excitation. These results validate the effectiveness of the experimental system. In Fig. 5(f) in the main text, we add defects by fixing the edge nodes at the zigzag boundary, from which the boundary condition is entirely changed from a zigzag shape to a bearded boundary.

\begin{figure}[htbp]
\begin{center}
\centerline{\includegraphics[width=0.8\linewidth]{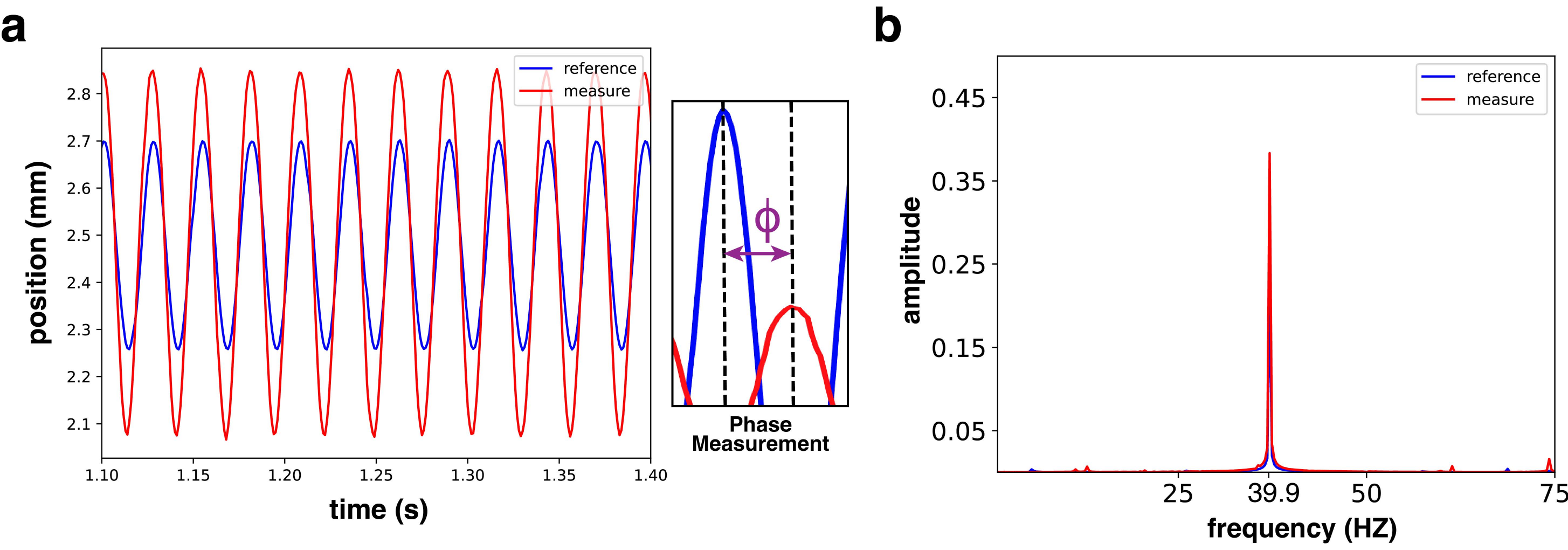}}
\begin{flushleft}\label{s7}FIG. S7: Schematic of experimental measurements. (a) left panel is measurement of time-zone vibrational curves of one lattice site and one reference site under excitation frequency 39.9HZ; Right panel is the phase measurement schematic. (b) Frequency spectrum after Fourier transform, where the peaks locates at 39.9HZ. \end{flushleft}
\vspace{-0cm}
\end{center}
\end{figure}

\subsection{\label{sec:level7}VII. Extension in other systems}

The approach proposed in this work to use ESP to induce topological states in valley architectures possesses significant scalability. This procedure is effectively reducible to a three-step protocol. First, choose a valley lattice material with Dirac cones (e.g., the hexagonal lattice), where a specific symmetry (such as sublattice symmetry) is broken to lift the Dirac degeneracy and open a band gap. Second, the equivalent term corresponding to the on-site potential within the system is identified; instances include the stiffness of grounding springs in mechanical lattices. Finally, by spatially engineering this on-site potential—such as through the quadratic and Gaussian distributions in the main text or the morphologies discussed in SI Section IV —one can calculate and observe the topological evolution of the in-gap bands.

Both extensions in the main text follow this procedure. For graphene-like electronic systems, adjusting the unit cell potential modifies the on-site term and effective mass, while in phononic crystals, this is achieved by controlling cavity volume or wall impedance. Both implementations successfully induce valley-protected topological states in the band gap. We next demonstrate the extension of this method to LC oscillatory circuits.

\begin{figure}[htbp]
\begin{center}
\centerline{\includegraphics[width=0.8\linewidth]{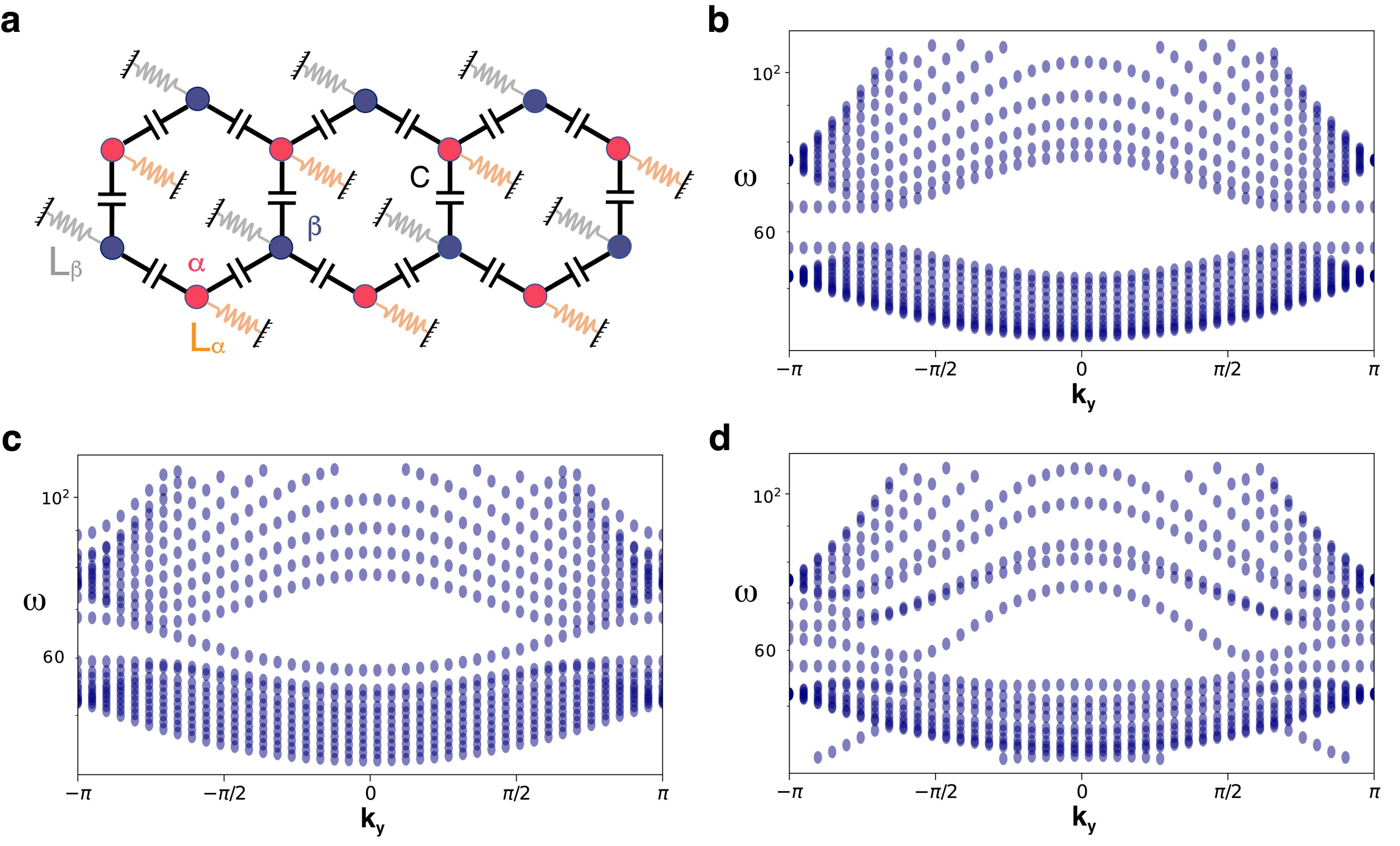}}
\begin{flushleft}\label{s8}FIG. S8: Extension to LC circuit system. (a) Schematic of the LC circuit hexagonal lattice composed of capacitance $C$ and inductance $L_{\alpha/\beta}$. (b-d) exhibit the spectra of quasi-1D LC circuit chains which is finite along x-axis and infinite in y-axis. (b) plots the original chain with $L_\alpha, L_\beta$ difference but without ESP. (c) illustrates the quadratic ESP at critical strength. (d) exhibits the Gaussian ESP in the middle at critical strength. \end{flushleft}
\vspace{-0cm}
\end{center}
\end{figure}

We consider a one-dimensional periodic LC circuit chain. In this configuration, each node $n$ is grounded via an inductor $L$ and coupled to its nearest neighbors ($n-1$ and $n+1$) through capacitors $C$. This setup effectively maps the spring constant $k$ to the coupling capacitance $C$, and the mass term $m$ to the grounding inductance $L$ as in the mechanical network.

Applying Kirchhoff's Current Law at node $n$ with voltage $V_n(t)$, the conservation of charge yields:
\begin{equation*}
    C \frac{d}{dt}(V_n - V_{n-1}) + C \frac{d}{dt}(V_n - V_{n+1}) + I_{L,n} = 0
\end{equation*}
where $I_{L,n}$ is the current flowing through the grounding inductor, satisfying the constitutive relation $V_n = L \frac{d I_{L,n}}{dt}$. Differentiating equation with respect to time $t$ eliminates the integral term, leading to the second-order differential equation of the circuit:
\begin{equation*}
    C \frac{d^2}{dt^2}(2V_n - V_{n-1} - V_{n+1}) + \frac{1}{L} V_n = 0
\end{equation*}
which exhibits a clear duality with Newton's equations of motion for a lattice. Assuming a harmonic ansatz $V_n(t) = v_n e^{i\omega t}$. Substituting this into the equation of motion gives:
\begin{equation*}
    -\omega^2 C (2v_n - v_{n-1} - v_{n+1}) + \frac{1}{L} v_n = 0
\end{equation*}
Rearranging the terms, we obtain the stationary Schrödinger-like equation for the circuit lattice:
\begin{equation*}
    2C v_n - C (v_{n-1} + v_{n+1}) = \frac{1}{\omega^2 L} v_n
\end{equation*}
Comparing it with the standard tight-binding model $\epsilon_0 \psi_n - t(\psi_{n-1} + \psi_{n+1}) = E \psi_n$, we identify the following correspondences with the mechanical system in the main text:
\begin{itemize}
    \item {Coupling (Spring $k$):} The hopping amplitude is determined by the capacitance, $t \leftrightarrow C$.
    \item {On-site Term:} The on-site potential is related to the connectivity, $\epsilon_0 \leftrightarrow 2C$.
    \item {Eigenvalue (Energy):} The spectral parameter depends on the frequency and inductance, $E \leftrightarrow \frac{1}{\omega^2 L}$.
\end{itemize}
Thus, by tuning the capacitor $C$, we directly control the inter-site coupling strength, while the inductor $L$ sets the resonant frequency scale of the system. Building on this foundation, we extended our study to an LC circuit system to verify the universality of the ESP-induced topological states (Fig. S8). In (a) we show the circuit scheme, couplings consist of identical capacitors $C$, while on-site potentials are modeled by grounded inductors ($L_\alpha, L_\beta$). We realized the equivalent ESP field by spatially modulating the inductance of the sublattices. Here, the sublattice nodes correspond to circuit nodes, and the oscillation signal is defined by the node's voltage. The band structures in Figs. S8(b-d) confirm our findings: (b) depicts the initial band gap, while (c) and (d) illustrate topological states emerging under quadratic and Gaussian ESPs, consistent with the main text.

We note two differences relative to the mechanical system: (1) The spectrum is spectrally inverted due to the governing equations, and (2) the eigenvalue spacing differs, necessitating an exponential scale. Nevertheless, as observed in the graphene-like and acoustic examples, the ESP successfully induces the expected topological states. While the specific governing equations introduce minor spectral variations, they do not compromise the fundamental role of the ESP. This methodology is equally extensible to photonic crystals, which are not discussed here due to space constraints.

\subsection{\label{sec:level7}VIII. Topological lithography using Gaussian ESP in electron systems}

\begin{figure}[htbp]
\begin{center}
\centerline{\includegraphics[width=1.0\linewidth]{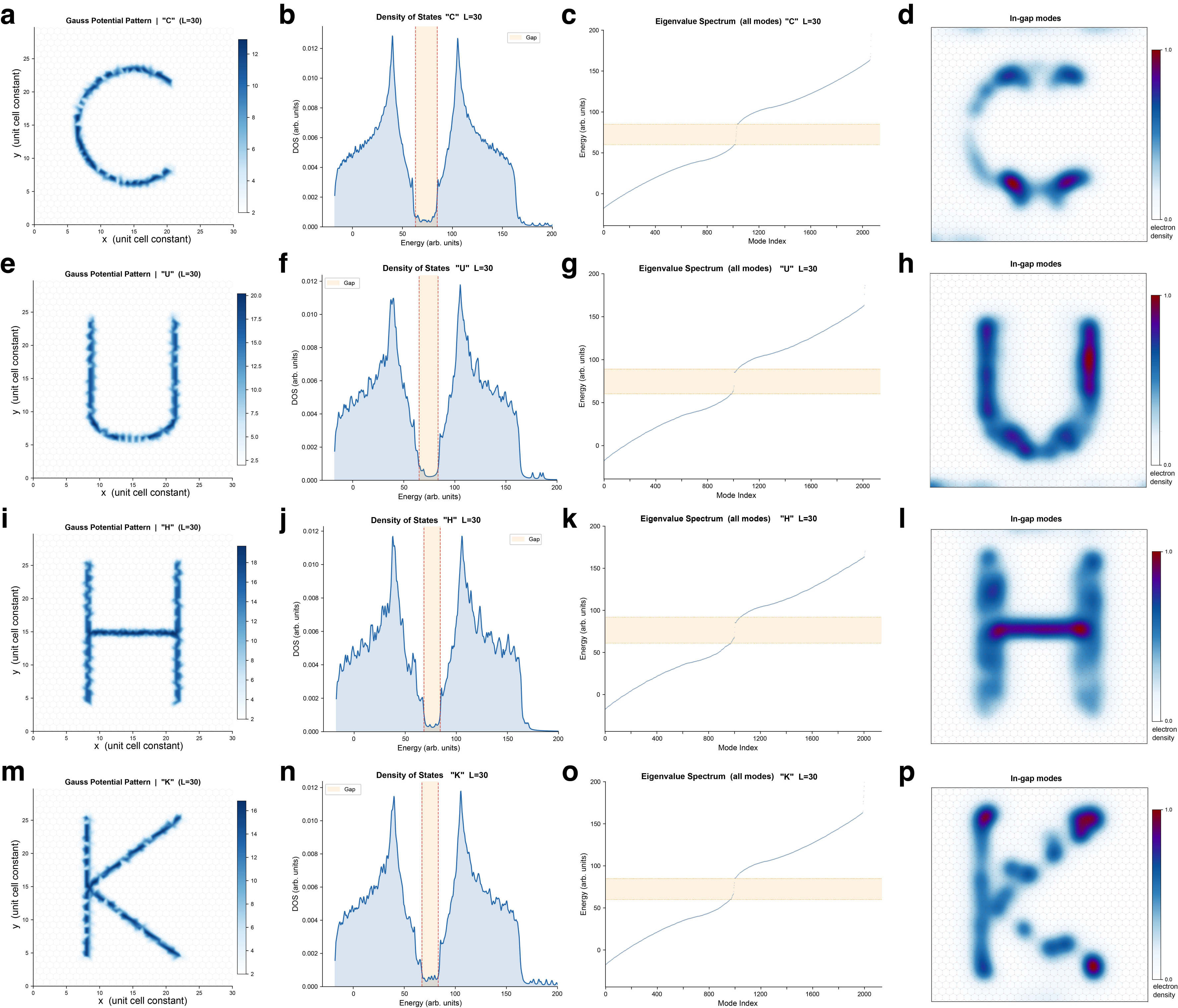}}
\begin{flushleft}\label{s8}FIG. S9: Topological lithography in electron systems. The first column shows the distribution of the scalar potential field applied on the lattice area. The second and third columns exhibit the density of states (DOS) and eigenvalue spectra, respectively. The gap areas are indicated by orange color. The last column illustrates the simulated in-gap mode distributions of the corresponding lithography patterns. \end{flushleft}
\vspace{-0cm}
\end{center}
\end{figure}

The main text demonstrates the concept of topological lithography realized through a Gaussian-profile ESP in an electronic lattice system. Here we provide a more detailed account of the construction and the resulting phenomenology, as illustrated in Fig. S9. The finite hexagonal lattices possess parameters identical to the main text: hopping term 30 units, original onsite potentials $p_A=30$ units and $p_B=42.67$ units, and ESP strength $\gamma=16$. The target patterned characters are lithographically encoded into the lattice via a Gaussian ESP whose amplitude is held constant along the stroke direction of each character while decaying as a Gaussian profile in the transverse direction, thereby producing a smooth and well-defined potential landscape as exhibited in Fig. S9(a,e,i,m).

For each encoded pattern, the density of states (DOS) spectrum is computed by solving the corresponding electronic Hamiltonian, and the results are presented in Fig. S9(b,f,j,n), where the orange-shaded regions denote the band gap. In all cases, a topological edge state threading continuously through the band gap is clearly identified, confirming that the ESP-induced topologically protected in-gap modes. The corresponding eigenvalue spectra are shown in Fig. S9(c,g,k,o), with the band gap again highlighted in orange, and are fully consistent with the DOS analysis.

The spatial profiles of the in-gap modes, obtained by direct numerical diagonalization of the finite system, are displayed in Fig. S9(d,h,l,p). The mode distributions are found to follow the prescribed ESP patterns with high fidelity, confirming that the topological edge states are spatially guided by the lithography landscape. Minor perturbative defects are visible at certain locations along the pattern boundaries; these are attributed to finite-size boundary effects inherent to the truncated lattice geometry rather than to any intrinsic limitation of the topological lithography mechanism. Overall, the simulated topological lithography results are in excellent agreement with theoretical expectations, providing compelling evidence for the efficacy and precision of the proposed paradigm.

We note that in realistic electronic and phononic lattice platforms, the requisite ESP can be introduced through a variety of experimentally accessible means, including scanning probe tips, substrate-induced strain fields, and externally applied stress distributions. The ability to pattern topological edge states with spatial resolution determined by the ESP profile, combined with the broad range of available implementation strategies, endows topological lithography with substantial practical utility for the design of reconfigurable topological circuitry and functional quantum material architectures.